\documentclass[aps,prd,nofootinbib,onecolumn,superscriptaddress,showpacs,showkeys]{revtex4}

\usepackage{graphicx}
\usepackage{epsfig,latexsym,amssymb}
\usepackage{amsmath}
\usepackage{bm}
\usepackage{simpler-wick}
\usetikzlibrary{decorations.markings}

\makeatother

\usepackage{slashed}
\usepackage[compat=1.0.0]{tikz-feynman}
\usepackage{comment}

\begin{document}

\title{Perturbative structure of two- and four-point functions
of color charge in a non-Gaussian small-$x$ action}

\author{Andre V. Giannini}
\affiliation{Instituto de F\'isica Gleb Wataghin, Universidade Estadual de Campinas,
R. S\'ergio Buarque de Holanda, 777, 13083-859, Campinas, Brazil}

\affiliation{
Instituto de F\'{i}sica, Universidade de S\~ao Paulo,%
Rua do Mat\~ao 1371,  05508-090, S\~ao Paulo-SP, Brazil}

\author{Yasushi Nara}
\affiliation{Akita International University, Yuwa, Akita-city 010-1292, Japan}

\begin{abstract}
We compute the perturbative expansion of the two- and four-point functions
of color charges in the Color Glass Condensate framework considering the 
quartic correction to the McLerran-Venugopalan (MV) model of Gaussian 
color charge fluctuations.
Expressions for these correlators in the perturbative expansion 
for small and large non-Gaussian color charge fluctuations
are derived for arbitrary orders in perturbation theory.
We explicitly show that the perturbative series does not converge at 
higher orders as expected. We apply the Borel-Pad\'e resummation method 
to our problem to construct a convergent series.
It is shown that the fully non-perturbative solution can be 
described by the Borel-Pad\'e approximants constructed from the first few terms 
of the perturbative series for small non-Gaussian fluctuations.
\end{abstract}

\maketitle

\section{Introduction}

The color glass condensate (CGC) effective theory~\cite{CGC.review.new}
has been successively used in understanding initial particle production processes
at high energy hadronic collisions and the initial phase of collisions of
heavy nuclei.
The Mclerran-Venugopalan (MV) model~\cite{CGC.Raju.McLerran}
of color charge fluctuations in nuclei assumes a Gaussian weight function, which is believed to be a good approximation for large systems. One expects that non-Gaussian corrections to the MV model may become important for describing color charge fluctuations in small collision systems. These corrections may also be relevant to the dijet correlations. In particular, the recent observation of a ``ridge'' in two-particle correlations from high-multiplicity $pp$ and $pA$ collisions~\cite{ridgePP,ridgePA}
motivates us to consider the non-Gaussian effects.
The ridge for Au + Au collisions~\cite{ridgeAA} is believed to be due to
collective effects and the signature of the formation of quark-gluon plasma,
which can be described by hydrodynamic evolution~\cite{Dusling:2015gta,Nagle:2018nvi}.
An outstanding question remains: whether the ridge effect observed in collisions involving small systems is due to collective effects or initial state dynamics~\cite{Dusling:2015gta,Nagle:2018nvi,Schlichting:2016sqo}.

Non-Gaussian corrections for the weight function have been
derived up to the fourth-order in the color charges~\cite{Jeon:2004rk,Dumitru:2011zz}.
In Ref.~\cite{Dumitru:2011zz}, the two- and four-point functions of the color charges
are computed at leading order in the regime where the quartic term is assumed to be a small perturbation, and showed that infrared behavior of the leading connected two-particle production diagram is different from the case of a quadratic action.
We discussed the properties of the non-perturbative solution in Ref.~\cite{Giannini:2020xme} for arbitrarily large non-Gaussian fluctuations.
The non-Gaussian action could be employed for generating initial conditions for the JIMWLK
evolution equation~\cite{JalilianMarian:1996xn,JalilianMarian:1997jx,JalilianMarian:1997gr,JalilianMarian:1997dw,JalilianMarian:1998cb,Kovner:1999bj,Kovner:2000pt,Iancu:2000hn,Iancu:2001ad,Ferreiro:2001qy}; in particular, it would provide corrections to the JIMWLK evolution equation for the four-point function of color charges not considered in~\cite{Dumitru:2010mv}. 

In this paper, we extend the work of Ref.~\cite{Dumitru:2011zz} by computing 
these color charge correlators at higher orders in perturbation theory. 
We explicitly compute diagrams up to next-to-next-leading-order (NNLO).
Computing each diagram is helpful to understand the structure of the theory. 
However, when going to several orders beyond the leading one, it may not be useful anymore to see the complicated graphs. Instead, we propose a general method to compute two- and four-point functions at any order in the perturbation.
It is well known that
the perturbative expansion in quantum mechanics and quantum field theory often has zero radius of convergence~\cite{Dyson:1952tj,Marino:2012zq,Kazakov:1980rd,Flory:2012nk}. 
The asymptotic character of perturbation theory is also suggested by 
the fact that the number of Feynman diagrams at order $n$ typically grows factorially.
In general, perturbative solution yields extremely good accuracy, 
e.g. quantum electrodynamics, but 
it will diverge at the terms of order of the inverse of coupling constant. 
We shall show that our perturbative series is also divergent,
and the optimal order for truncating the perturbative series depends on
the values of coupling constant.

The Borel algorithm in perturbation theory is a ressumation method to construct a convergent perturbative series, see e.g.~\cite{Zinn,Itzykson}.
Based on the Borel method, several resummation methods are proposed~\cite{Caliceti:2007ra,Mera:2018qte}.
We will apply a widely used one for our problem: the Borel-Pad\'e resummation method~\cite{Caliceti:2007ra}.
In this method, a Pad\'e approximant replaces the Borel-transformed series expansion
by a rational function whose numerator and denominator are
chosen so that its power series expansion agrees with the original power series
up to the term whose degree is equal to the sum of the degrees of the numerator and
the denominator of the rational function.
We shall show that almost exact solutions can be obtained by taking only the first few terms of the divergent series.

This paper is organized as follows.
After describing the setup of our problem in Section~\ref{sec:setup},
we work out perturbative expressions for the limit of small non-Gaussian 
fluctuations and compare its results to the full non-perturbative calculation in Section~\ref{sec:kappa}. A formulation of the perturbative series in the limit of large non-Gaussian fluctuations is given in Sec.~\ref{sec:z}.
In Sec.~\ref{sec:BorelPade}, we show the results of the Borel-Pad\'e approximant for our problem.
The conclusion is given in Sec.~\ref{sec:conclusion}.

\section{Color charge averaging for small/large deviations from the MV model}
\label{sec:setup}

We consider the asymptotic expansion of the quartic correction to the MV model
\footnote{We note that we absorbed a factor 3 
	into the definition of $\kappa$: $\kappa = \kappa_4/3$, as
	compared to~\cite{Giannini:2020xme}. We also inverted 
	the notation for the coupling in the MV model and the quadratic 
	term in the non-Gaussian action to make comparisons 
	with~\cite{Dumitru:2011zz} easier: 
	$\mu$ in this paper is the coupling of the quadratic term in the 
	non-Gaussian action while $\bar{\mu}$ is the renormalized color 
	charge appearing in the MV model.}~\cite{Dumitru:2011zz}
\begin{equation}\label{eq:largeZperturbation}
\langle \mathcal{O}[\rho]\rangle
\equiv\frac{\int\mathcal{D}\rho\, \mathcal{O}[\rho]\,e^{-S_G-\int d^2w \rho^4_w/\kappa}}
{\int\mathcal{D}\rho\, e^{-S_G - \int d^2w \rho^4_w/\kappa}}
=\frac{\int\mathcal{D}\rho\, \mathcal{O}[\rho]\,e^{-S_G}
\sum_{k=0}^\infty \frac{1}{k!}(-\frac{1}{\kappa}\int d^2w\, \rho^4_w)^k}
       {\int\mathcal{D}\rho\, e^{-S_G}
\sum_{k=0}^\infty \frac{1}{k!}(-\frac{1}{\kappa}\int d^2w\, \rho^4_w)^k}\,,
\end{equation}
where $S_G=\int d^2x\,\rho_x^2/2\mu^2$ denotes the Gaussian action.
On the other hand, in the regime of large non-Gaussian fluctuations, 
one may consider the expansion of the action as:
\begin{equation}
\int\mathcal{D}\rho_x\,\exp\left[-\frac{Z\rho_x^2}{2\bar\mu^2}-\frac{\rho_x^4}{\kappa}\right]
=\int\mathcal{D}\rho_x\, \sum_{k=0}^\infty\frac{1}{k!}\left(-\frac{Z\rho_x^2}{2\bar\mu^2}\right)^k
\exp\left[-\frac{\rho_x^4}{\kappa}\right]\,,
\end{equation}
where $Z=\bar\mu^2/\mu^2$ is the renormalization factor, and 
$\bar\mu$ is the coefficient of the two-point function in 
the MV model, {\it i}.{\it e}., the renormalized color charge:
\begin{equation}
\langle \rho^a(x)\rho^b(y)\rangle=
\langle \rho_x^a\rho_y^b\rangle=
\delta^{ab}
\delta(x-y)
\bar\mu^2\,
\label{eq:renom}
\end{equation}

Following \cite{Dumitru:2011zz,Giannini:2020xme}, we compute the functional 
integral in lattice regularization, assuming a square lattice with $N_s\times N_s$ 
sites of length $a$. In the case of a local operator, which is the one 
we consider here, the color charge average in the SU($N_c$) theory can be written as:
\begin{equation}
\langle \mathcal{O}_r\rangle = \frac{\int\, dr\,r^{N_c^2-2}\,\mathcal{O}_r\,e^{-W_r}}
                         {\int dr\,r^{N_c^2-2}\,e^{-W_r}}\,,
\end{equation}
where $r^2 = \sum_{a=1}^{N_c^2-1}\rho_{x}^a \rho_{x}^a$ and
\begin{equation}
W_r  = \frac{a^2\,r^2}{2\,\mu^2} + \frac{a^2\,r^4}{\kappa}\,.
 \label{eq:wr}
\end{equation}

In the next section, we compute the two- and four-point function of color charges in the 
limits where {\it i}) Eq. (\ref{eq:wr}) is dominated by its quadritic 
term, and  {\it ii}) Eq. (\ref{eq:wr}) is dominated by its quartic
term, corresponding to the regimes of small and large non-Gaussian 
fluctuations, respectively, and compare the result from different 
orders in perturbation to the non-perturbative result. 

\section{Perturbation theory in the limit of small non-Gaussian fluctuations} 
\label{sec:kappa}

A leading order (LO) calculation in $1/\kappa$ for two- and four-point function of color charges has already been presented
in~\cite{Dumitru:2011zz}. We start by summarizing their 
result, then extend calculations to next-to-leading 
order (NLO) and next-to-next-to-leading order (NNLO). 
Then, we present the structure of these correlators
at N$^n$LO order in perturbation theory.

\subsection{Leading order}

We summarize the LO result for the two- and four-point function~\cite{Dumitru:2011zz}.
The color factors of the basic diagrams contributing at this order 
are shown in Fig.~\ref{fig:base_color}, where $x$, $y$, $u$, and $v$ 
represent coordinates in position space and $a$, $b$, $c$, and $d$ 
represent color indexes. We suppress color indexes in other 
figures to avoid cluttered diagrams.
\begin{figure}[htb]
\begin{equation}
\begin{tikzpicture}[baseline=(e)]
\begin{feynman}[inline=(e)]
\vertex (a){\((y,b)\)};
\vertex[below=2 cm of a ](b){\((x,a)\)};
\vertex[above=1cm of b](e);
\vertex[right=1cm of e](h);
\diagram*{
    (a)--(b)
};
\end{feynman}
\end{tikzpicture}
=\frac{\mu^2}{a^2}\delta^{ab}\delta_{xy}
\begin{tikzpicture}[baseline=-\the\dimexpr\fontdimen22\textfont2\relax]
\begin{feynman}[inline=(e)]
\vertex (e);
\vertex[above=0.5cm of e](h);
\vertex[below=0.5cm of e](g);
\node at (0.3,0.0) {$w$};
\diagram*{
    (e)--[out=45,in=0,min distance=0.5cm](h),
    (e)--[out=135,in=180,min distance=0.5cm](h),
    (e)--[out=-45,in=0,min distance=0.5cm](g),
    (e)--[out=-135,in=180,min distance=0.5cm](g)
};
\end{feynman}
\end{tikzpicture}
=(N_c^4-1) N_s\frac{\mu^4}{a^2}
~~~
\begin{tikzpicture}[baseline=(e)]
\begin{feynman}[inline=(e)]
\vertex (a){\((y,b)\)};
\vertex[below=2 cm of a ](b){\((x,a)\)};
\vertex[above=1cm of b](e);
\vertex[right=1cm of e](h);
\node at (0.4,-1.0) {$w$};
\diagram*{
    (e)--[out=45,in=90,min distance=0.5cm](h),
    (e)--[out=-45,in=-90,min distance=0.5cm](h), (a)--(b)
};
\end{feynman}
\end{tikzpicture}
=4(N_c^2+1)\frac{\mu^6}{a^4}\delta^{ab}\delta_{xy}
~~~\begin{tikzpicture}[baseline=(x)]
\begin{feynman}[inline=(x)]
\vertex (a){\((x,a)\)};
\vertex[below=0.75 cm of a ](x);
\vertex[right=1.5 cm of a ](b){\((v,d)\)};
\vertex[below=1.5 cm of b] (c){\((y,b)\)};
\vertex[left=1.5 cm of c](d){\((u,c)\)};
\node at (0.7,-1.0) {$w$};
\path (a)--(c) coordinate[pos=0.25] (e);
\coordinate (h) at ($(e)+(-{1/sqrt(2)},-{1/sqrt(2)})$);
\diagram*{
    (d)--[draw=white,double=black,very thick](b)
    ,(c)--(a)
};
\end{feynman}
\end{tikzpicture}
=8\frac{\mu^8}{a^6}\delta^{ab}\delta^{cd}\delta_{xy}\delta_{xu}\delta_{xv}
\end{equation}
\caption{The color factor for diagrams contributing at LO for the
two- and four-point function of color charges.}
\label{fig:base_color}
\end{figure}

From Eq.~(\ref{eq:largeZperturbation}), the two-point 
function at the order $1/\kappa$
\begin{equation}
\langle\rho_x^a\,\rho_y^b\rangle
=\frac{\int\mathcal{D}\rho\, \rho_x^a\,\rho_y^b\,e^{-S_G}\,
	[1-\frac{1}{\kappa}\,\int d^2w\, \rho^4_w]}
       {\int\mathcal{D}\rho\, e^{-S_G}[1-\frac{1}{\kappa}\,\int d^2w\, \rho^4_w]}\,,
\end{equation}
can be expressed diagrammatically as: 
\begin{eqnarray}
\langle \rho^a_x\,\rho^b_y\rangle&=&
\left(
\begin{tikzpicture}[baseline=(e)]
\begin{feynman}[inline=(e)]
\vertex[right=0.4 cm of a ](b){\(y\)};
\vertex[below=1.0 cm of b] (c){\(x\)};
\diagram*{
    (c)--(b),
};
\end{feynman}
\end{tikzpicture}
-\frac{1}{\kappa}\left[
\begin{tikzpicture}[baseline=(e)]
\begin{feynman}[inline=(e)]
\vertex[right=0.4 cm of a ](b){\(y\)};
\vertex[below=1 cm of b] (c){\(x\)};
\vertex[above=0.5 cm of c](c2);
\vertex[right=0.4cm of c2](e);
\vertex[above=0.3cm of e](h);
\vertex[below=0.3cm of e](g);
\node at (1.1,-0.5) {$w$};
\diagram*{
    (c)--(b),
    (e)--[out=45,in=0,min distance=0.2cm](h),
    (e)--[out=135,in=180,min distance=0.2cm](h),
    (e)--[out=-45,in=0,min distance=0.2cm](g),
    (e)--[out=-135,in=180,min distance=0.2cm](g)
};
\end{feynman}
\end{tikzpicture}
+
\begin{tikzpicture}[baseline=(h)]
\begin{feynman}[inline=(h)]
\vertex[right=0.7 cm of a ](b){\(y\)};
\vertex[below=1 cm of b] (c){\(x\)};
\vertex[above=0.5cm of c](e);
\vertex[right=0.5cm of e](h);
\node at (1.0,-0.5) {$w$};
\diagram*{
    (e)--[out=45,in=90,min distance=0.5cm](h),
    (e)--[out=-45,in=-90,min distance=0.5cm](h), (c)--(b)
};
\end{feynman}
\end{tikzpicture}
\right]
\right)
\left(
1-\frac{1}{\kappa}
\begin{tikzpicture}[baseline=(e)]
\begin{feynman}[inline=(e)]
\vertex (a);
\vertex[below=0.5cm of a](e);
\vertex[above=0.5cm of e](h);
\vertex[below=0.5cm of e](g);
\node at (0.3,-0.5) {$w$};
\diagram*{
    (e)--[out=45,in=0,min distance=0.2cm](h),
    (e)--[out=135,in=180,min distance=0.2cm](h),
    (e)--[out=-45,in=0,min distance=0.2cm](g),
    (e)--[out=-135,in=180,min distance=0.2cm](g)
};
\end{feynman}
\end{tikzpicture}
\right)^{-1}\\
&=&
\left(
\begin{tikzpicture}[baseline=(e)]
\begin{feynman}[inline=(e)]
\vertex (a){\(y\)};
\vertex[below=1.0 cm of a](d){\(x\)};
\vertex[below=0.5cm of  a](e);
\diagram*{
    (d)--(a)
};
\end{feynman}
\end{tikzpicture}
-\frac{1}{\kappa}
\begin{tikzpicture}[baseline=(h)]
\begin{feynman}[inline=(h)]
\vertex (a){\(y\)};
\vertex[below=1.0 cm of a](d){\(x\)};
\vertex[above=0.5cm of d](e);
\vertex[right=0.5cm of e](h);
\node at (0.3,-0.5) {$w$};
\diagram*{
    (d)--(a), (e)--[out=45,in=90,min distance=0.5cm](h),
    (e)--[out=-45,in=-90,min distance=0.5cm](h)
};
\end{feynman}
\end{tikzpicture}
\right)
=\mu^2\frac{\delta^{ab}\delta_{xy}}{a^2}
\left(
1-4(N_c^2+1)\frac{\mu^4}{\kappa a^2}
\right)\,.
\end{eqnarray}
Given that the two-point function of color charges must 
be the same in both the MV model and its non-Gaussian 
extension, we identify the renormalized the average 
charge squared at order ${\mathcal O}(1/\kappa)$ as:
\begin{equation}
\bar\mu^2=\mu^2\left[1-4\frac{\mu^4}{\kappa a^2}(N_c^2+1)\right]
\label{eq:renomCharge}
\end{equation}
Thus, the two-point function reads
\begin{equation}
\langle \rho^a_x\rho^b_y\rangle=
\frac{\delta^{ab}\delta_{xy}}{a^2}\bar\mu^2\,.
\end{equation}
The four-point function at leading order in $1/\kappa$ 
evaluates to:
\begin{eqnarray}
\langle \rho^a_x\rho^b_y\rho_u^c\rho_v^d\rangle&=&
\left(
\begin{tikzpicture}[baseline=(e)]
\begin{feynman}[inline=(e)]
\vertex (a){\(y\)};
\vertex[right=0.4 cm of a ](b){\(v\)};
\vertex[below=1.0 cm of b] (c){\(u\)};
\vertex[left=0.4 cm of c](d){\(x\)};
\vertex[above=0.5 cm of c](c2);
\vertex[right=0.3cm of c2](e);
\diagram*{
    (d)--(a), (c)--(b),
};
\end{feynman}
\end{tikzpicture}
-\frac{1}{\kappa}\left[
\begin{tikzpicture}[baseline=(e)]
\begin{feynman}[inline=(e)]
\vertex (a){\(y\)};
\vertex[right=0.4 cm of a ](b){\(v\)};
\vertex[below=1 cm of b] (c){\(u\)};
\vertex[left=0.4 cm of c](d){\(x\)};
\vertex[above=0.5 cm of c](c2);
\vertex[right=0.4cm of c2](e);
\vertex[above=0.3cm of e](h);
\vertex[below=0.3cm of e](g);
\node at (1.1,-0.5) {$w$};
\diagram*{
    (d)--(a), (c)--(b),
    (e)--[out=45,in=0,min distance=0.2cm](h),
    (e)--[out=135,in=180,min distance=0.2cm](h),
    (e)--[out=-45,in=0,min distance=0.2cm](g),
    (e)--[out=-135,in=180,min distance=0.2cm](g)
};
\end{feynman}
\end{tikzpicture}
+
\begin{tikzpicture}[baseline=(h)]
\begin{feynman}[inline=(h)]
\vertex (a){\(y\)};
\vertex[right=0.7 cm of a ](b){\(v\)};
\vertex[below=1 cm of b] (c){\(u\)};
\vertex[left=0.7 cm of c](d){\(x\)};
\vertex[above=0.5cm of c](e);
\vertex[right=0.5cm of e](h);
\node at (0.4,-0.5) {$w$};
\diagram*{
    (d)--(a), (e)--[out=45,in=90,min distance=0.5cm](h),
    (e)--[out=-45,in=-90,min distance=0.5cm](h), (c)--(b)
};
\end{feynman}
\end{tikzpicture}
+
\begin{tikzpicture}[baseline=(e)]
\begin{feynman}[inline=(e)]
\vertex (a){\(x\)};
\vertex[right=1 cm of a ](b){\(v\)};
\vertex[below=1 cm of b] (c){\(y\)};
\vertex[left=1 cm of c](d){\(u\)};
\node at (0.7,-0.5) {$w$};
\path (a)--(c) coordinate[pos=0.5] (e);
\diagram*{
    (d)--[draw=white,double=black,very thick](b)
    ,(c)--(a)
};
\end{feynman}
\end{tikzpicture}
\right]
\right)
\left(
1-\frac{1}{\kappa}
\begin{tikzpicture}[baseline=(e)]
\begin{feynman}[inline=(e)]
\vertex (a);
\vertex[below=0.5cm of a](e);
\vertex[above=0.5cm of e](h);
\vertex[below=0.5cm of e](g);
\node at (0.3,-0.5) {$w$};
\diagram*{
    (e)--[out=45,in=0,min distance=0.2cm](h),
    (e)--[out=135,in=180,min distance=0.2cm](h),
    (e)--[out=-45,in=0,min distance=0.2cm](g),
    (e)--[out=-135,in=180,min distance=0.2cm](g)
};
\end{feynman}
\end{tikzpicture}
\right)^{-1}\\
&=&
\left(
\begin{tikzpicture}[baseline=(e)]
\begin{feynman}[inline=(e)]
\vertex (a){\(y\)};
\vertex[right=0.4 cm of a ](b){\(v\)};
\vertex[below=1.0 cm of b] (c){\(u\)};
\vertex[left=0.4 cm of c](d){\(x\)};
\vertex[above=0.5 cm of c](c2);
\vertex[right=0.3cm of c2](e);
\diagram*{
    (d)--(a), (c)--(b),
};
\end{feynman}
\end{tikzpicture}
-\frac{1}{\kappa}\left[
\begin{tikzpicture}[baseline=(h)]
\begin{feynman}[inline=(h)]
\vertex (a){\(y\)};
\vertex[right=0.7 cm of a ](b){\(v\)};
\vertex[below=1 cm of b] (c){\(u\)};
\vertex[left=0.7 cm of c](d){\(x\)};
\vertex[above=0.5cm of c](e);
\vertex[right=0.5cm of e](h);
\node at (0.4,-0.5) {$w$};
\diagram*{
    (d)--(a), (e)--[out=45,in=90,min distance=0.5cm](h),
    (e)--[out=-45,in=-90,min distance=0.5cm](h), (c)--(b)
};
\end{feynman}
\end{tikzpicture}
~+
\begin{tikzpicture}[baseline=(e)]
\begin{feynman}[inline=(e)]
\vertex (a){\(x\)};
\vertex[right=1 cm of a ](b){\(v\)};
\vertex[below=1 cm of b] (c){\(y\)};
\vertex[left=1 cm of c](d){\(u\)};
\node at (0.7,-0.5) {$w$};
\path (a)--(c) coordinate[pos=0.5] (e);
\diagram*{
    (d)--[draw=white,double=black,very thick](b)
    ,(c)--(a)
};
\end{feynman}
\end{tikzpicture}
\right]
\right)\\
&=&\frac{\mu^4}{a^4}(
\delta^{ab}\delta_{xy}\delta^{cd}\delta_{uv}
+\delta^{ac}\delta_{xu}\delta^{bd}\delta_{yv}
+\delta^{ad}\delta_{xv}\delta^{bc}\delta_{yu}
)\left[
1 - 8\frac{\mu^4}{\kappa a^2}(N_c^2+1)
\right] \nonumber\\
&-&8\frac{\mu^8}{\kappa a^6}
(\delta^{ab}\delta^{cd}
+\delta^{ac}\delta^{bd}
+\delta^{ad}\delta^{bc})\delta_{xy}\delta_{xu}\delta_{uv}\,.
\end{eqnarray}

Eq.~(\ref{eq:renomCharge}) implies 
\begin{equation}
\bar\mu^4=\mu^4\left[1-8\frac{\mu^4}{\kappa a^2}(N_c^2+1)\right]\,,
\end{equation}
at leading order in $1/\kappa$.
The term $\mu^8/\kappa$ is replaced by $\mu^8\to\bar\mu^8$ 
as any higher-order contribution in $1/\kappa$ is discarded 
at LO.
Thus, the four-point function at LO in $1/\kappa$ becomes
\begin{eqnarray}
\langle \rho^a_x\rho^b_y\rho_u^c\rho_v^d\rangle&=&
\frac{\bar\mu^4}{a^4}(
\delta^{ab}\delta_{xy}\delta^{cd}\delta_{uv}
+\delta^{ac}\delta_{xu}\delta^{bd}\delta_{yv}
+\delta^{ad}\delta_{xv}\delta^{bc}\delta_{yu}
)
-8\frac{\bar\mu^8}{\kappa a^6}
(\delta^{ab}\delta^{cd}
+\delta^{ac}\delta^{bd}
+\delta^{ad}\delta^{bc})\delta_{xy}\delta_{xu}\delta_{uv}\,.
\end{eqnarray}

\subsection{The next-to-leading order}\label{sec:NLO_largeZ}

We now compute the two- and four-point functions 
using Eq. (\ref{eq:largeZperturbation}) at 
NLO in $1/\kappa$. 
The two-point function reads:
\begin{eqnarray}
\langle \rho^a_x\rho^b_y\rangle&=&
\left(
\begin{tikzpicture}[baseline=(e)]
\begin{feynman}[inline=(e)]
\vertex (a){\(y\)};
\vertex[below=1.0 cm of a](d){\(x\)};
\vertex[below=0.5cm of  a](e);
\diagram*{
    (d)--(a)
};
\end{feynman}
\end{tikzpicture}
-\frac{1}{\kappa}
\begin{tikzpicture}[baseline=(h)]
\begin{feynman}[inline=(h)]
\vertex (a){\(y\)};
\vertex[below=1.0 cm of a](d){\(x\)};
\vertex[above=0.5cm of d](e);
\vertex[right=0.5cm of e](h);
\node at (0.3,-0.5) {$w$};
\diagram*{
    (d)--(a), (e)--[out=45,in=90,min distance=0.5cm](h),
    (e)--[out=-45,in=-90,min distance=0.5cm](h)
};
\end{feynman}
\end{tikzpicture}
+\frac{1}{2\kappa^2}\left[
\begin{tikzpicture}[baseline=(c)]
\begin{feynman}[inline=(c)]
\vertex (a){\(y\)};
\vertex[below=0.75cm of a](c);
\vertex[below=1.5 cm of a](d){\(x\)};
\vertex[above=0.5cm of d](e);
\vertex[right=0.5cm of e](h);
\vertex[above=1.0cm of d](f);
\vertex[right=0.5cm of f](i);
\node at (0.2,-1.0) {$w$};
\node at (0.2,-0.5) {$z$};
\diagram*{
    (d)--(a),
    (e)--[out=30,in=90,min distance=0.3cm](h),
    (e)--[out=-30,in=-90,min distance=0.3cm](h),
    (f)--[out=30,in=90,min distance=0.3cm](i),
    (f)--[out=-30,in=-90,min distance=0.3cm](i)
};
\end{feynman}
\end{tikzpicture}
~+
\begin{tikzpicture}[baseline=(h)]
\begin{feynman}[inline=(h)]
\vertex (a){\(y\)};
\vertex[below=1.5 cm of a](d){\(x\)};
\vertex[below=0.75cm of a](e);
\vertex[right=0.4cm of e](f);
\vertex[right=0.4cm of e](h);
\vertex[right=0.4cm of f](i);
\node at (0.2,-0.75) {$z$};
\node at (0.6,-0.75) {$w$};
\diagram*{
    (d)--(a),
    (e)--[out=45,in=90,min distance=0.4cm](h),
    (e)--[out=-45,in=-90,min distance=0.4cm](h),
    (f)--[out=45,in=90,min distance=0.4cm](i),
    (f)--[out=-45,in=-90,min distance=0.4cm](i)
};
\end{feynman}
\end{tikzpicture}
+
\begin{tikzpicture}[baseline=(c)]
\begin{feynman}[inline=(c)]
\vertex (a){\(y\)};
\vertex[below=1.5 cm of a ](b){\(x\)};
\vertex[below=0.4cm of a](e);
\vertex[below=1.1cm of a](f);
\vertex[below=0.75cm of a](c);
\node at (0.2,-1.2) {$w$};
\node at (0.2,-0.35) {$z$};
\diagram*{
    (b)--(a),
   (f)--[out=-180,in=180,min distance=0.3cm](e),
   (f)--[out=0,in=0,min distance=0.3cm](e)
};
\end{feynman}
\end{tikzpicture}
\right]
\right)\\
&=&\mu^2\frac{\delta^{ab}\delta_{xy}}{a^2}
\left[
1-4\frac{\mu^4}{\kappa a^2}(N_c^2+1)
+\frac{\mu^8}{2\kappa^2a^4}
\left(
 32(N_c^2+1)^2+32(N_c^2+1)^2+32\times2(N_c^2+1)
\right)
\right]\\
&=&\mu^2\frac{\delta^{ab}\delta_{xy}}{a^2}
\left[
1-4\frac{\mu^4}{\kappa a^2}(N_c^2+1)
+32\frac{\mu^8}{\kappa^2a^4}
(N_c^2+1)(N_c^2+2)
\right]\,.
\end{eqnarray}
Thus, the color charge squared is renormalized as
\begin{equation}
\bar\mu^2 \equiv
\mu^2 
\left[
1-4\frac{\mu^4}{\kappa a^2}(N_c^2+1)
+32\frac{\mu^8}{\kappa^2a^4}
(N_c^2+1)(N_c^2+2)
\right]\,.
\end{equation}
For the four-point function
there are six connected diagrams that contribute 
at NLO, whose color factors are:
\begin{equation}\label{eq:diagrams_NLO_part1}
\begin{tikzpicture}[baseline=(h)]
\begin{feynman}[inline=(h)]
\vertex (a){\(x\)};
\vertex[right=2 cm of a ](b){\(v\)};
\vertex[below=2 cm of b] (c){\(y\)};
\vertex[left=2 cm of c](d){\(u\)};
\path (a)--(c) coordinate[pos=0.25] (e);
\coordinate (h) at ($(e)+(-{1/sqrt(2)},-{1/sqrt(2)})$);
\node at (0.4,-0.8) {$z$};
\node at (1.4,-1.0) {$w$};
\diagram*{
        (e)--[out=-180,in=135,min distance=0.5cm](h),
    (e)--[out=-90,in=-45,min distance=0.5cm](h), (c)--(a),
    (d)--[draw=white,double=black,very thick](b)
};
\end{feynman}
\end{tikzpicture}
=16\times16(N_c^2+1)
~~
\begin{tikzpicture}[baseline=(v)]
\begin{feynman}[inline=(v)]
\vertex (a){\(x\)};
\vertex[right=1.0 cm of a ](b){\(y\)};
\vertex[below=0.7 cm of a] (x);
\vertex[right=0.5 cm of x] (e);
\vertex[below=0.5 cm of e] (v);
\vertex[below=1.0 cm of e] (f);
\vertex[below=0.7 cm of f] (y);
\vertex[left=0.3 cm of y] (c){\(u\)};
\vertex[right=0.3 cm of y] (d){\(v\)};
\node at (0.5,-1.5) {$z$};
\node at (0.5,-0.9) {$w$};
\diagram*{
   (a)--(e),(e)--(b),(c)--(f),(d)--(f),
   (f)--[out=-180,in=180,min distance=0.5cm](e),
   (f)--[out=0,in=0,min distance=0.5cm](e)
};
\end{feynman}
\end{tikzpicture}
{\hskip -0.2cm = 16\times4(N_c^2+7)}
{\hskip 0.3cm}
\begin{tikzpicture}[baseline=(h)]
\begin{feynman}[inline=(h)]
\vertex (a){\(y\)};
\vertex[right=0.7 cm of a ](b){\(v\)};
\vertex[below=2 cm of b] (c){\(u\)};
\vertex[left=0.7 cm of c](d){\(x\)};
\vertex[above=0.5cm of c](e);
\vertex[below=1.0cm of b](h);
\vertex[above=1.5cm of c](f);
\vertex[right=0.5cm of f](i);
\node at (0.85,-1.65) {$w$};
\node at (0.85,-0.35) {$z$};
\diagram*{
    (d)--(a),
    (c)--(b),
   (f)--[out=-180,in=180,min distance=0.3cm](e),
   (f)--[out=0,in=0,min distance=0.3cm](e)
};
\end{feynman}
\end{tikzpicture}
= 8\times16(N_c^2+1)
\end{equation}
\begin{eqnarray}
\begin{tikzpicture}[baseline=(h)]
\begin{feynman}[inline=(h)]
\vertex (a){\(y\)};
\vertex[right=0.7 cm of a ](b){\(v\)};
\vertex[below=2 cm of b] (c){\(u\)};
\vertex[left=0.7 cm of c](d){\(x\)};
\vertex[above=1cm of c](e);
\vertex[right=0.5cm of e](h);
\vertex[above=1cm of d](f);
\vertex[left=0.5cm of f](i);
\node at (1.0,-1.0) {$w$};
\node at (-0.3,-1.0) {$z$};
\diagram*{
    (d)--(a),
    (e)--[out=45,in=90,min distance=0.5cm](h),
    (e)--[out=-45,in=-90,min distance=0.5cm](h),
    (c)--(b),
    (f)--[out=135,in=90,min distance=0.5cm](i),
    (f)--[out=-135,in=-90,min distance=0.5cm](i)
};
\end{feynman}
\end{tikzpicture}
~~~=[4(N_c^2+1)]^2\times2
\begin{tikzpicture}[baseline=(x)]
\begin{feynman}[inline=(x)]
\vertex (a){\(y\)};
\vertex[right=0.7 cm of a ](b){\(v\)};
\vertex[below=2 cm of b] (c){\(u\)};
\vertex[left=0.7 cm of c](d){\(x\)};
\vertex[above=0.5cm of c](e);
\vertex[right=0.5cm of e](h);
\vertex[above=1.5cm of c](f);
\vertex[right=0.5cm of f](i);
\vertex[below=1.0 cm of a ](x);
\node at (1.0,-1.5) {$w$};
\node at (1.0,-0.5) {$z$};
\diagram*{
    (d)--(a),
    (e)--[out=45,in=90,min distance=0.5cm](h),
    (e)--[out=-45,in=-90,min distance=0.5cm](h),
    (c)--(b),
    (f)--[out=45,in=90,min distance=0.5cm](i),
    (f)--[out=-45,in=-90,min distance=0.5cm](i)
};
\end{feynman}
\end{tikzpicture}
=16(N_c^2 + 1)^2\times2\times2
\begin{tikzpicture}[baseline=(h)]
\begin{feynman}[inline=(h)]
\vertex (a){\(y\)};
\vertex[right=0.7 cm of a ](b){\(v\)};
\vertex[below=2 cm of b] (c){\(u\)};
\vertex[left=0.7 cm of c](d){\(x\)};
\vertex[above=1.0cm of c](e);
\vertex[right=0.5cm of e](f);
\vertex[right=0.5cm of e](h);
\vertex[right=0.5cm of f](i);
\node at (1.4,-1.0) {$z$};
\node at (0.95,-1.0) {$w$};
\diagram*{
    (d)--(a),
    (e)--[out=45,in=90,min distance=0.5cm](h),
    (e)--[out=-45,in=-90,min distance=0.5cm](h),
    (c)--(b),
    (f)--[out=45,in=90,min distance=0.5cm](i),
    (f)--[out=-45,in=-90,min distance=0.5cm](i)
};
\end{feynman}
\end{tikzpicture}
&=&16(N_c^2 + 1)^2\times2\times2\,.
\end{eqnarray}

Let us define the four-point function as
\begin{equation}
\langle \rho_x^a\rho_y^b\rho_u^c\rho_v^d\rangle
= 
(\delta^{ab}\delta_{xy}\delta^{cd}\delta_{uv}
+\delta^{ac}\delta_{xu}\delta^{bd}\delta_{yv}
+\delta^{ad}\delta_{xv}\delta^{bc}\delta_{yu})
 A_\mathrm{dis} + 
 (\delta^{ab}\delta^{cd}
+\delta^{ac}\delta^{bd}
+\delta^{ad}\delta^{bc})\delta_{xy}\delta_{xu}\delta_{uv}A_\mathrm{con}\,,
\end{equation}
where the disconnected (MV) part, $A_\mathrm{dis}$, must be the square 
of the two-point function (at the given order in perturbation theory) 
as seen in the computation at leading order presented in the previous 
section:
\begin{eqnarray}
A_\mathrm{dis}
&=&
\frac{\mu^4}{a^4}\left[
  1-8\frac{\mu^4}{\kappa a^2}(N_c^2+1)
 + 64\frac{\mu^8}{\kappa^2a^4}(N_c^2+1)(N_c^2+2)
 + 16\frac{\mu^8}{\kappa^2a^4}(N_c^2+1)^2
  \right]
= \frac{\bar\mu^4}{a^4}\,.
\end{eqnarray}
The connected part is given by
\begin{equation}
A_\mathrm{con}
= \frac{\mu^4}{a^4}
 \left[
   -8\frac{\mu^{4}}{\kappa a^2}
   \left(1-\frac{16\mu^4}{\kappa a^2}(N_c^2+1)\right)
  +32\frac{\mu^8}{\kappa^2a^4}(N_c^2+7)
  \right]\,.
\end{equation}
The first term is renormalized by the left-most diagram 
in Eq. (\ref{eq:diagrams_NLO_part1})
and we can replace $\mu\to\bar\mu$
in the second term as it is already a 
term of order ${\mathcal{O}(1/\kappa^2)}$:
\begin{equation}
A_\mathrm{con} = \frac{\bar\mu^4}{a^4}
 \left[
   -8\frac{\bar\mu^{4}}{\kappa a^2}
  +32\frac{\bar\mu^8}{\kappa^2a^4}(N_c^2+7)
  \right]\,.
\end{equation}

An explicit computation of each diagram that 
contributes for the two- and four-point function of color charge at NNLO 
($\mathcal{O}(1/\kappa^3)$) is presented in the
Appendix \ref{appendix:nnlo}.

%

We shall work out the structure of these correlators for an arbitrary 
order in perturbation theory in the next section.
For this purpose,
it would be helpful to show Eq.~(\ref{eq:twopointNLO})
and Eq.~(\ref{eq:fourpointNLO}) in a slightly different way, which can be 
extend to the evaluation of the renormalized coefficients at N$^n$LO.
The non-vanishing part of the two- and four-point 
function before and after re-normalization may be 
expressed as
\begin{eqnarray}
 \langle \rho^a_x\rho^a_x \rangle &=&
 (N_c^2-1)\frac{\mu^2}{a^2}\left[ 1 - c\,x(1 + a_1x )
 \right]
 = (N_c^2-1)\frac{\mu^2}{a^2}\left[ 1 - c\, y(1 + c_1y)
 \right]
 = (N_c^2-1)\frac{\bar\mu^2}{a^2}
 \label{eq:twopointNLO},
 \\
 \langle \rho^a_x\rho^a_x\rho^b_x\rho^b_x\rangle &=&
 (N_c^4-1)\frac{\mu^4}{a^4}\left( 1 +a_1 x  + a_2 x^2
	    \right)
 = (N_c^4-1)\frac{\bar\mu^4}{a^4}\left( 1 +c_1 y + c_2 y^2
	    \right),
 \label{eq:fourpointNLO}
\end{eqnarray}
where
\begin{equation}
c=4(N_c^2+1),\quad a_1=-8(N_c^2+2),\quad a_2=16\left(5N_c^4 + 24N_c^2 + 31\right),
\quad c_1= -8,\quad c_2= 32(N_c^2 + 7)\,,
\end{equation}
and $x=\mu^4/\kappa a^2$ and $y=\bar\mu^4/\kappa a^2$.
From Eq.(\ref{eq:twopointNLO}), the renomalization condition
is obtained as
\begin{equation}
 \mu^2\left[ 1 - c\,x(1 + a_1x ) \right] =\bar\mu^2
\end{equation}
from which we have
\begin{equation}
 x\left[ 1 - c\,x(1 + a_1x ) \right]^2 =y\,.
\end{equation}
Solving this equation for $x$ up to the next-to-leading order yields:
\begin{equation}
 x = y[1+2cy + (7c^2+2ca_1)y^2]\,.
\end{equation}
Then, the four-point function can be renormalized as
\begin{equation}
 x(1+a_1x + a_2x^2) = y[1+ (a_1+2c)y + (7c^2+6ca_1+a_2)y^2]\,.
\end{equation}
Thus, we also obtained $c_1 = a_1 + 2c=-8$ and $c_2=7c^2+6ca_1 + a_2=32(N_c^2+7)$.

It is worth recognizing that the two-point function has the same coefficients
as the four-point function: $c_1$, in the case of NLO.
In other words, the coefficients of the four-point function
are obtained by dividing the coefficients of the two-point 
function by $c=4(N_c^2+1)$ as expected by the cutting rule.
Starting from the vacuum graph at order $1/\kappa$, we note 
that the contribution 
for the two-point function can be obtained by cutting a loop from that 
graph in all possible ways, as indicated in Eq. (\ref{eq:cut_diagrams})
(a cut is indicated by the $\times$ symbol). 
The fully connected graph for the four-point function at order $1/\kappa$ can be obtained 
by cutting a loop in the resulting diagram for the two-point 
function, which has been shown in the case of 
scalar field theory with
$\phi^4$ self-interaction~\cite{phi4}:
\begin{equation}\label{eq:cut_diagrams}
\begin{tikzpicture}[baseline=-\the\dimexpr\fontdimen22\textfont2\relax]
\begin{feynman}[inline=(e)]
\vertex (e);
\vertex[above=0.5cm of e](h);
\vertex[below=0.5cm of e](g);
\node at (0.3,0.0) {$w$};
\node at (0.0,0.5) {$\times$};
\diagram*{
    (e)--[out=45,in=0,min distance=0.5cm](h),
    (e)--[out=135,in=180,min distance=0.5cm](h),
    (e)--[out=-45,in=0,min distance=0.5cm](g),
    (e)--[out=-135,in=180,min distance=0.5cm](g)
};
\end{feynman}
\end{tikzpicture}
~~
=
~~
\begin{tikzpicture}[baseline=(e)]
\begin{feynman}[inline=(e)]
\vertex (a){\(y\)};
\vertex[below=2 cm of a ](b){\(x\)};
\vertex[above=1cm of b](e);
\vertex[right=1cm of e](h);
\node at (0.4,-1.0) {$w$};
\node at (1.0,-1.0) {$\times$};
\diagram*{
    (e)--[out=45,in=90,min distance=0.5cm](h),
    (e)--[out=-45,in=-90,min distance=0.5cm](h), (a)--(b)
};
\end{feynman}
\end{tikzpicture}
~~
=
~~~\begin{tikzpicture}[baseline=(x)]
\begin{feynman}[inline=(x)]
\vertex (a){\(x\)};
\vertex[below=0.75 cm of a ](x);
\vertex[right=1.5 cm of a ](b){\(v\)};
\vertex[below=1.5 cm of b] (c){\(y\)};
\vertex[left=1.5 cm of c](d){\(u\)};
\node at (0.7,-1.0) {$w$};
\path (a)--(c) coordinate[pos=0.25] (e);
\coordinate (h) at ($(e)+(-{1/sqrt(2)},-{1/sqrt(2)})$);
\diagram*{
    (d)--[draw=white,double=black,very thick](b)
    ,(c)--(a)
};
\end{feynman}
\end{tikzpicture}\,.
\end{equation}
We note that scalar $\phi^4$ interaction has the same diagrams
as our problem in the perturbative expansion.
The corresponding color factor for the second diagram above may be 
obtained by dividing the vacuum graph by $(N_c^2-1)$,
then one may multiply it by the \textit{MV-part of} the two-point function:
$\mu^2\delta^{ab}\delta_{xy}/a^2$ to get the 
${\mathcal{O}(1/\kappa)}$ contribution to the two-point function of color charges.
The tadpole diagram always have the color factor $4(N_c^2+1)$.
This factor disappears after cutting the loop.
The ${\mathcal{O}(1/\kappa)}$ contribution for the four-point function of color 
charges can be obtained by dividing the color factor of the ${\mathcal{O}(1/\kappa)}$ 
contribution to the two-point function by $4(N_c^2+1)$ and then multiplying it 
by $\mu^2\delta^{ab}\delta_{xy}/a^2$. 

\begin{figure}[htb]
\includegraphics[width=8.0cm]{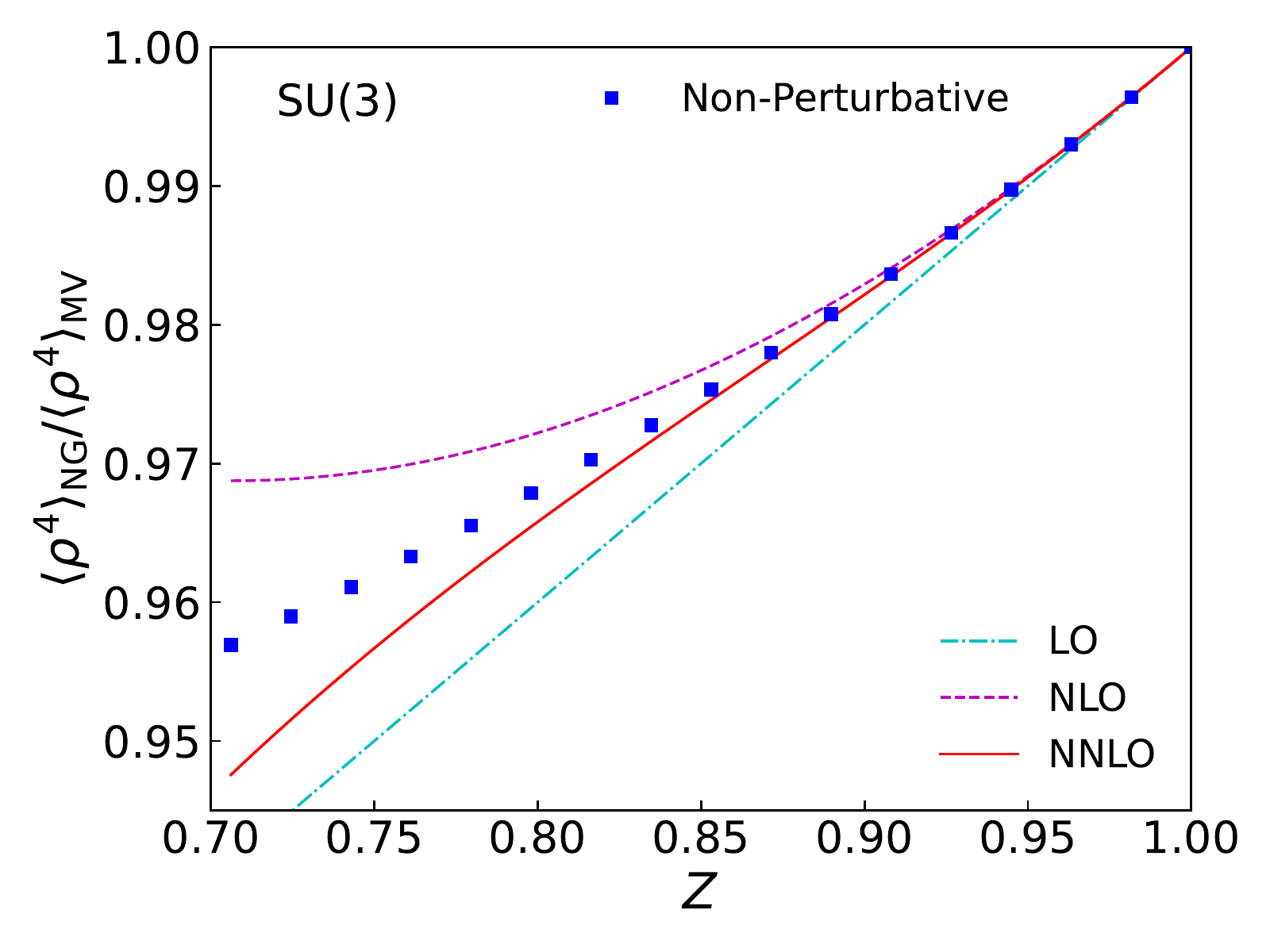}
\includegraphics[width=8.0cm]{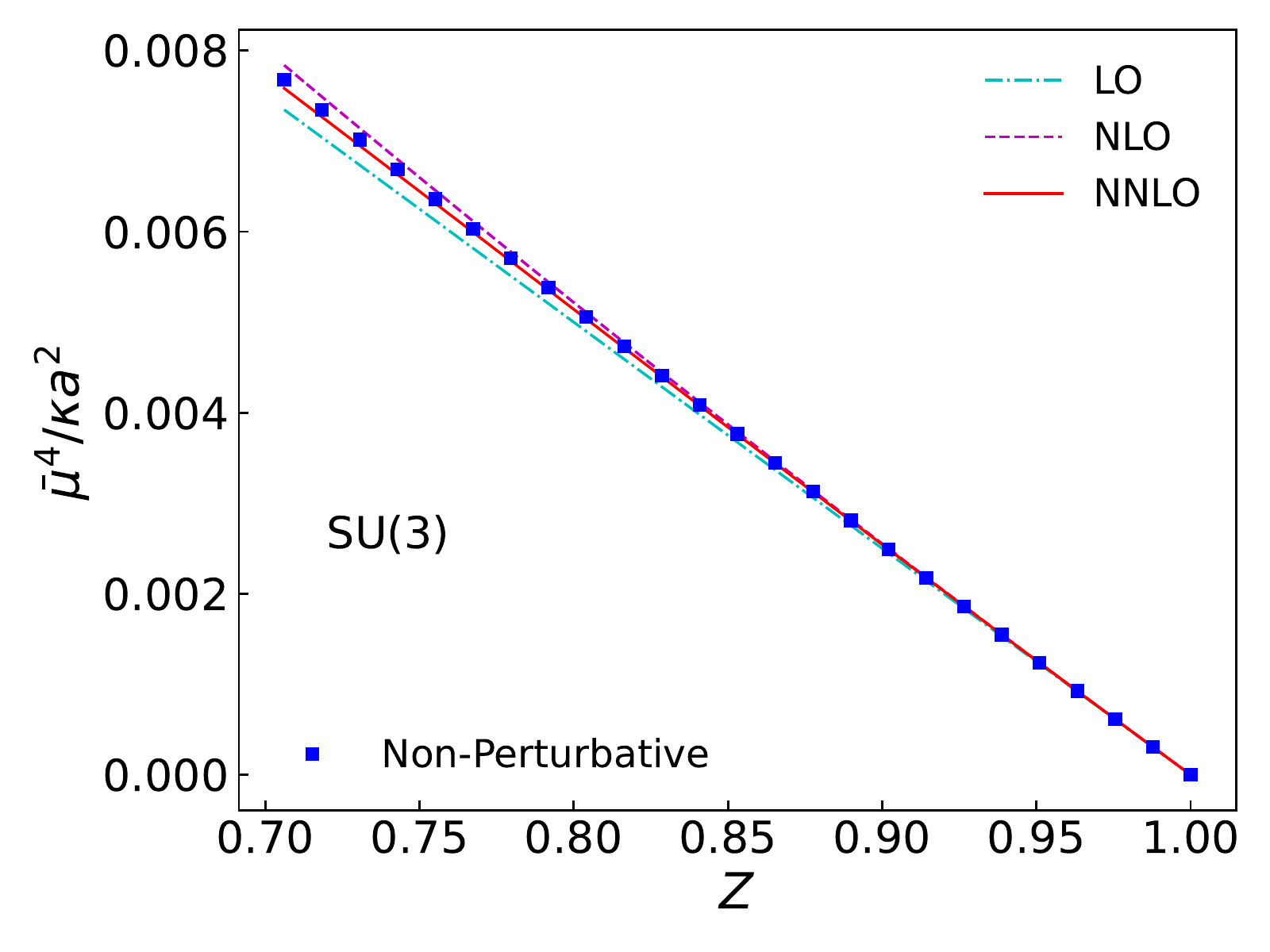}
\caption{The left panel shows the ratio of the four-point function as a function of $Z$
from LO, NLO, and NNLO calculations for SU(3).
The right panel shows the $Z$ dependence of  $y=\bar\mu^4/\kappa a^2$.
The full squares represent the result of the non-perturbative calculation.
}
\label{fig:ratioNNLO}
\end{figure}

In Fig.~\ref{fig:ratioNNLO}, we compare the LO, NLO, and NNLO results
with the non-perturbative result for the four point function (left panel) and the solution of the renomalization equation (right panel).
We obtain a better agreement as the perturbative order increases.
We see that the NNLO result is enough to reproduce the exact solution 
when the deviation from the Gaussian approximation is less than 10\%.
In the next section,
we will show the N$^n$LO results.

\subsection{The N$^n$LO order}

We consider a perturbative expansion in $1/\kappa$ at an arbitrary order.
To get a total sum of all diagrams, it is easier to work with the 
Gaussian integrals rather than by using the Wick contraction.
We define $N_n$ as
\begin{equation}\label{eq:Nn_largeZ}
N_n = \frac{\int \prod_a d\rho^a_x\, \rho_x^n\, e^{-S_G}}
 {\int \prod_a d\rho^a_x\, e^{-S_G}}\,,
\end{equation}
for which we have the recursive expression:
\begin{equation}\label{eq:NnGaussianIntegral}
	N_{n+2} = (N_c^2-1+n)\left(\frac{\mu}{a}\right)^2\,N_{n}, \qquad {\rm with}\quad N_0=1\,,
\end{equation}
for $n\geq 0$ and even. A derivation of Eq.~(\ref{eq:NnGaussianIntegral}) is presented in 
the Appendix \ref{appendix:GaussianIntegral}.
By using $N_n$, the two-point and four-point function at N$^n$LO can be expressed as:
\begin{eqnarray}
\langle \rho^a_{x}\rho^a_{x}\rangle
&=&\left(\sum_{k=0}^{n+1}\frac{1}{k!}\left(-\frac{a^2}{\kappa}\right)^kN_{4k+2}\right)
\left(\sum_{k=0}^{n+1}\frac{1}{k!}\left(-\frac{a^2}{\kappa}\right)^kN_{4k}\right)^{-1}
=(N_c^2-1) \frac{\mu^2}{a^2}
\left[
1-c \sum_{k=0}^n a_kx^{k+1}  
\right] \\ 
\langle \rho^a_{x}\rho^a_{x}\rho^b_{x}\rho^b_{x}\rangle
&=&\left(\sum_{k=0}^{n+1}\frac{1}{k!}\left(-\frac{a^2}{\kappa}\right)^kN_{4k+4}\right)
\left(\sum_{k=0}^{n+1}\frac{1}{k!}\left(-\frac{a^2}{\kappa}\right)^kN_{4k}\right)^{-1}
=(N_c^4-1)\frac{\mu^4}{a^4} \sum_{k=0}^{n+1}a_kx^k\,, 
\label{eq:4point}
\end{eqnarray}
where $x=\mu^4/\kappa a^2$.
Note that the four-point function has the same coefficients $a_k$ since 
it can be obtained by
cutting diagrams of two-point function at each order
as mentioned in previous sections.
The factor $c=4(N_c^2+1)$ is a contribution of a loop, which disappears 
after cutting the loop when calculating the four-point function.
The coefficients $a_k$ are listed in the Appendix~\ref{appendix:coefficients_ak_c_ck} up to ten orders in perturbation.

Imposing the condition that the two-point function must be matched 
with the MV model is written as:
\begin{equation}
Z\equiv\frac{\bar\mu^2}{\mu^2}= 1- c \sum_{k=0}^{n}a_k x^{k+1}\,,
\label{eq:z}
\end{equation}
where $\bar\mu^2$ is the renormalized color charge squared 
appearing in the two-point function in the MV model.
The renormalized expression is obtained by substituting
\begin{eqnarray}
\mu^2 &=& \bar\mu^2(1 - c x - c a_1x^2 \cdots)^{-1}\nonumber\\
&=& \bar\mu^2(1 + c x + ( c a_1 + c^2 ) x^2 
 + (c a_2 + 2 c^2 a_1 + c^3 ) x^3
 + ( c a_3 + 2 c^2  a_2 + c^2  a_1^2  + 3 c^3  a_1 + c^4 ) x^4
   +\cdots)
\label{eq:mubar2}
\end{eqnarray}
successively until all powers of $x$ are replaced by powers 
of $y=\bar\mu^4/\kappa a^2$ at a desired order
so that
we obtain the renomalization of Eq.~(\ref{eq:z}):
\begin{eqnarray}
  Z&=& 1 - c \sum_{k=0}^n c_k y^{k+1} \,.
\label{eq:rz}
\end{eqnarray}
The four-point function can be renormalized following the exact 
same procedure. After renormaliation, Eq.~(\ref{eq:4point})
may be written as:
\begin{eqnarray}
\langle \rho^a_{x}\rho^a_{x}\rho^b_{x}\rho^b_{x}\rangle
&=&(N_c^4-1)\frac{\bar\mu^4}{a^4} \sum_{k=0}^{n+1} c_k y^k \,.
\label{eq:r4point}
\end{eqnarray}
The coefficients $c_k$ can be found in the Appendix \ref{appendix:coefficients_ak_c_ck}.
The four-point function of color charges 
in continuum notation becomes
\begin{eqnarray}
\langle \rho_{x}^a\rho_{y}^b\rho_{u}^c\rho_{v}^d\rangle
&=& \bar\mu^4 \Bigg[
 \delta^{ab} \delta^{cd} \delta(x-y) \delta(u-v)
 \left(1 + \frac{\bar\mu^4}{\kappa}A_\mathrm{con}\,\delta(x-u)\right) \nonumber\\
&+&\delta^{ac} \delta^{bd} \delta(x-u) \delta(y-v)
 \left(1 + \frac{\bar\mu^4}{\kappa}A_\mathrm{con}\,\delta(x-y)\right) \nonumber\\
&+&\delta^{ad} \delta^{bc} \delta(x-v) \delta(y-u)
 \left(1 +\frac{\bar\mu^4}{\kappa} A_\mathrm{con}\,\delta(x-y)\right)
 \Bigg]\,,
\end{eqnarray}
where
\begin{equation}
A_\mathrm{con} = c_1 + c_2 y + c_3y^2 + c_4y^3 + \cdots + c_{n+1}y^{n}
\end{equation}
with $y$ determined by solving Eq.~(\ref{eq:rz}).

The ratio of the four-point function is given by
\begin{eqnarray}
R_n = \frac{\langle \rho_{x}^4\rangle}{\langle \rho_{x}^4\rangle_\mathrm{MV}}
&=& \sum_{k=0}^{n+1} c_k y^k 
= \frac{1-Z}{cy'}\,,
\label{eq:ratio1}
\end{eqnarray}
where $y'$ is a solution of Eq.~(\ref{eq:rz}) at order $y^{n+2}$.
These results suggests that we can sum divergent series:
\begin{equation}
    R_\infty = \sum_{k=0}^{\infty} c_k y^k 
= \frac{1-Z}{cy_\infty}\,,
\end{equation} 
providing that we know a exact solution $y_\infty$ for the renormalization equation. 


\begin{figure}[htb]
\includegraphics[width=8.0cm]{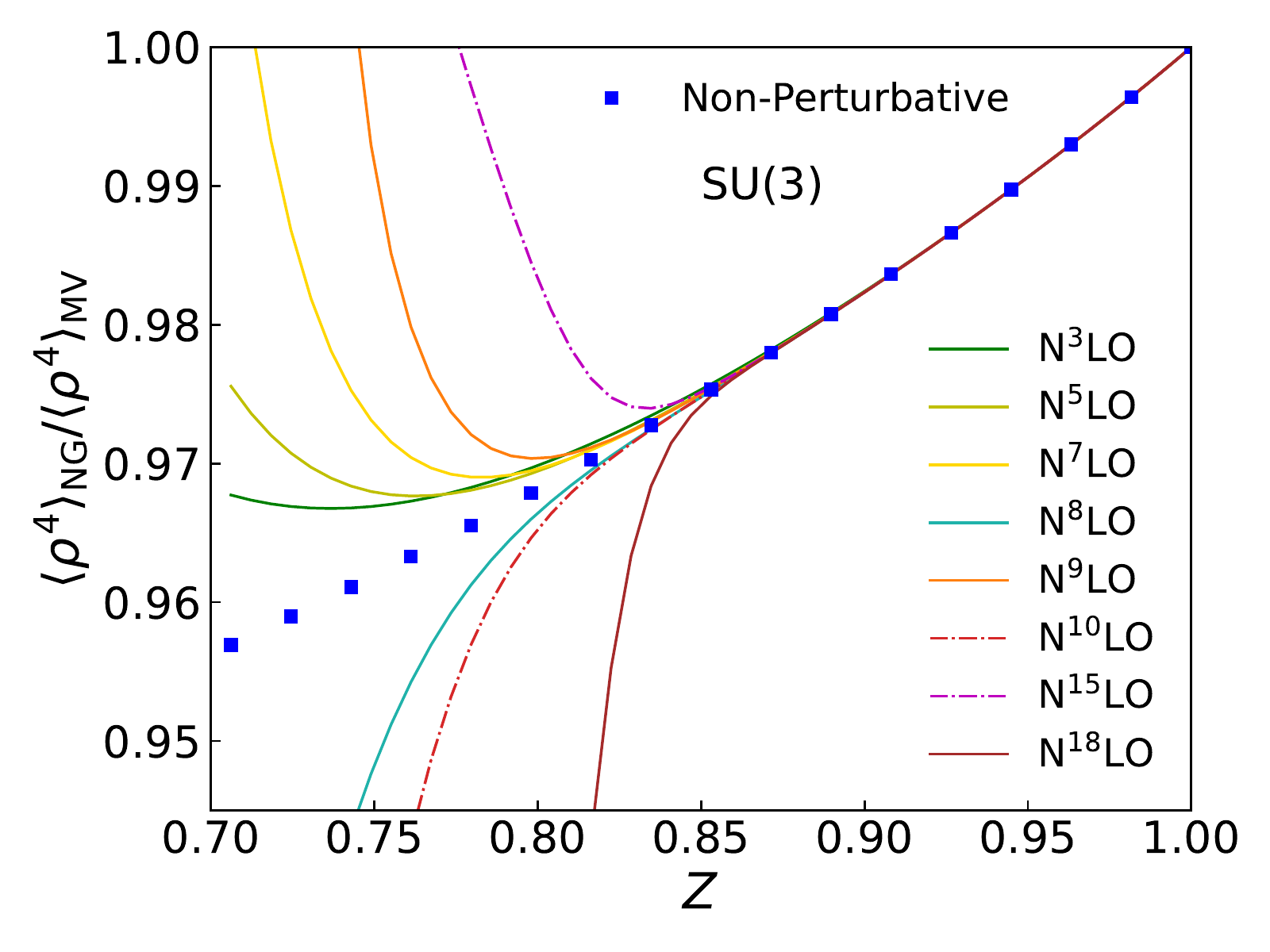}
\includegraphics[width=8.0cm]{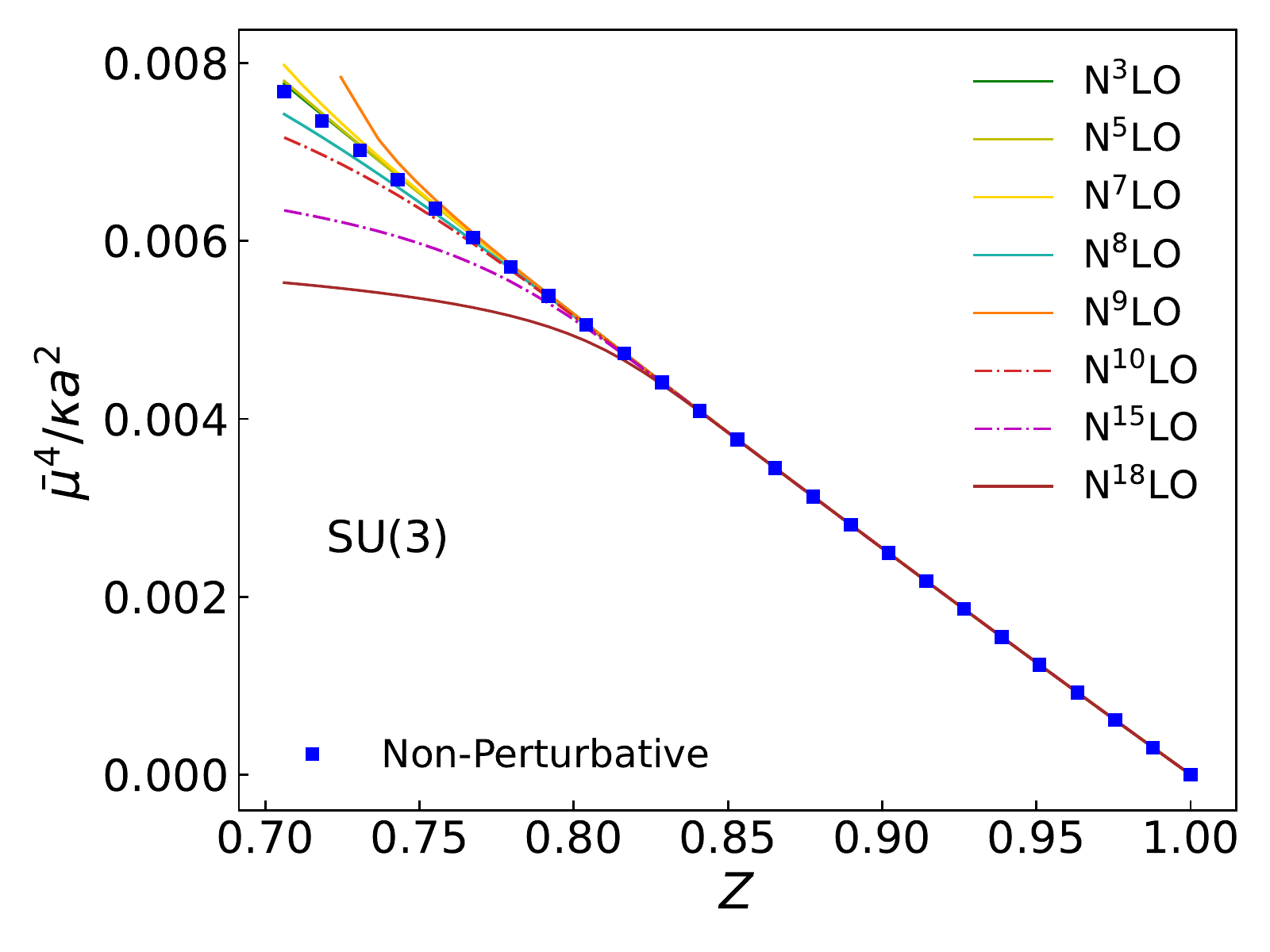}
\caption{Left panel: The ratio of the four-point function as a function of $Z$ for SU(3).
Right panel: The $Z$ dependence of $y=\bar\mu^4/\kappa a^2$ for different orders in perturbation theory. 
The full circles represent the result of the non-perturbative calculation.
}
\label{fig:ratio}
\end{figure}

We plot the result from Eq.~(\ref{eq:ratio1}) in Fig.~\ref{fig:ratio}
for SU(3) group as a function of $Z$ up to the 
eighteenth order in the expansion.
These ratios go up (down) for N$^{2n+1}$LO (N$^{2n}$LO).
Such a behaviour indicates that a large value is added or 
subtracted at each order. The perturbation in $1/\kappa$ works up to $Z\sim 0.85$, 
and perturbative solutions strongly disagree with the full solution.
The right panel shows the $Z$ dependence of $y=\bar\mu^4/\kappa a^2$. 
We can see that the perturbative solutions also deviate from the full solution  around
$Z=0.85$, which is the source of the divergence in the ratio of the four-point functions.

\begin{figure}[htb]
\includegraphics[width=8.0cm]{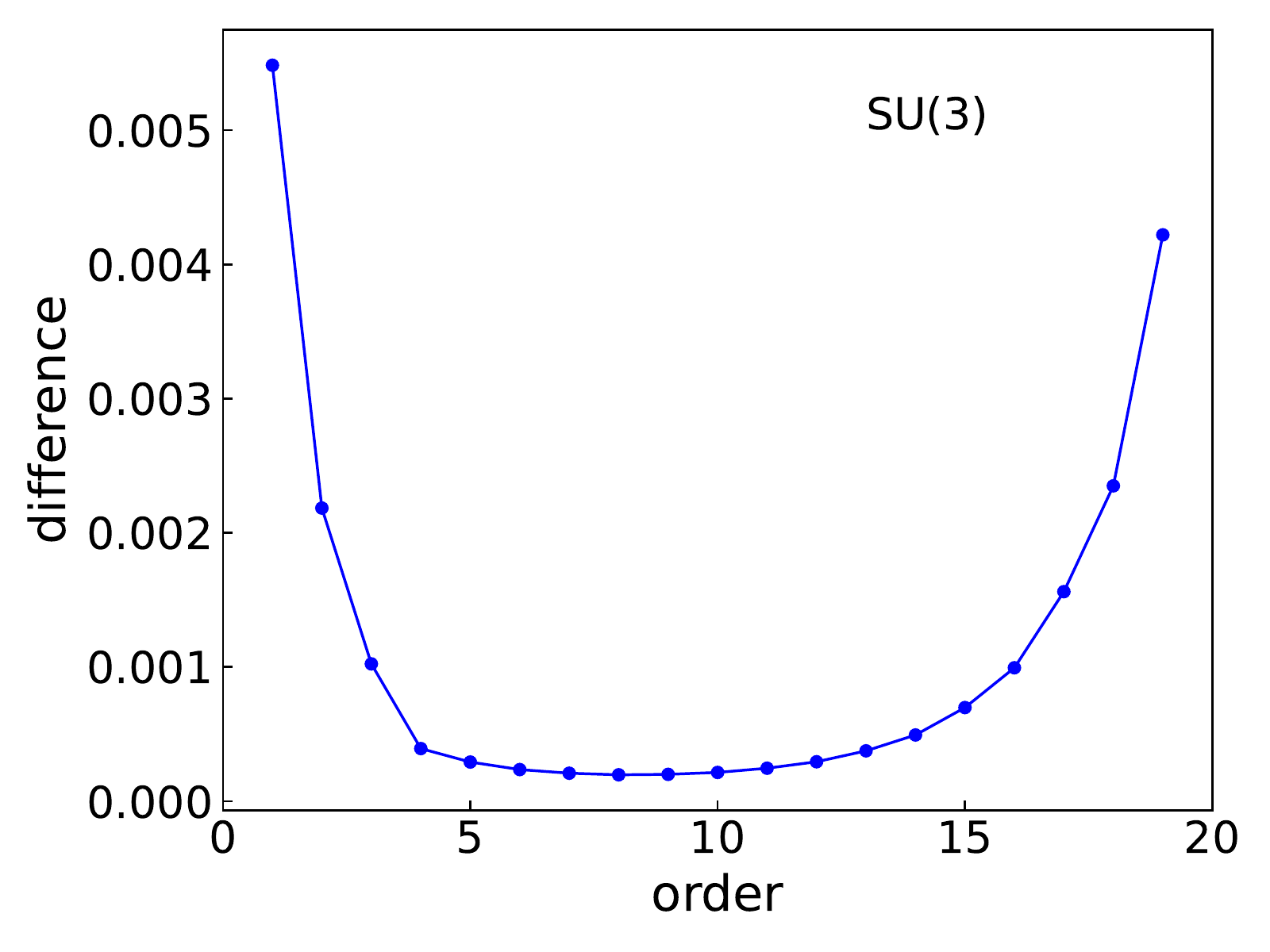}
\includegraphics[width=8.0cm]{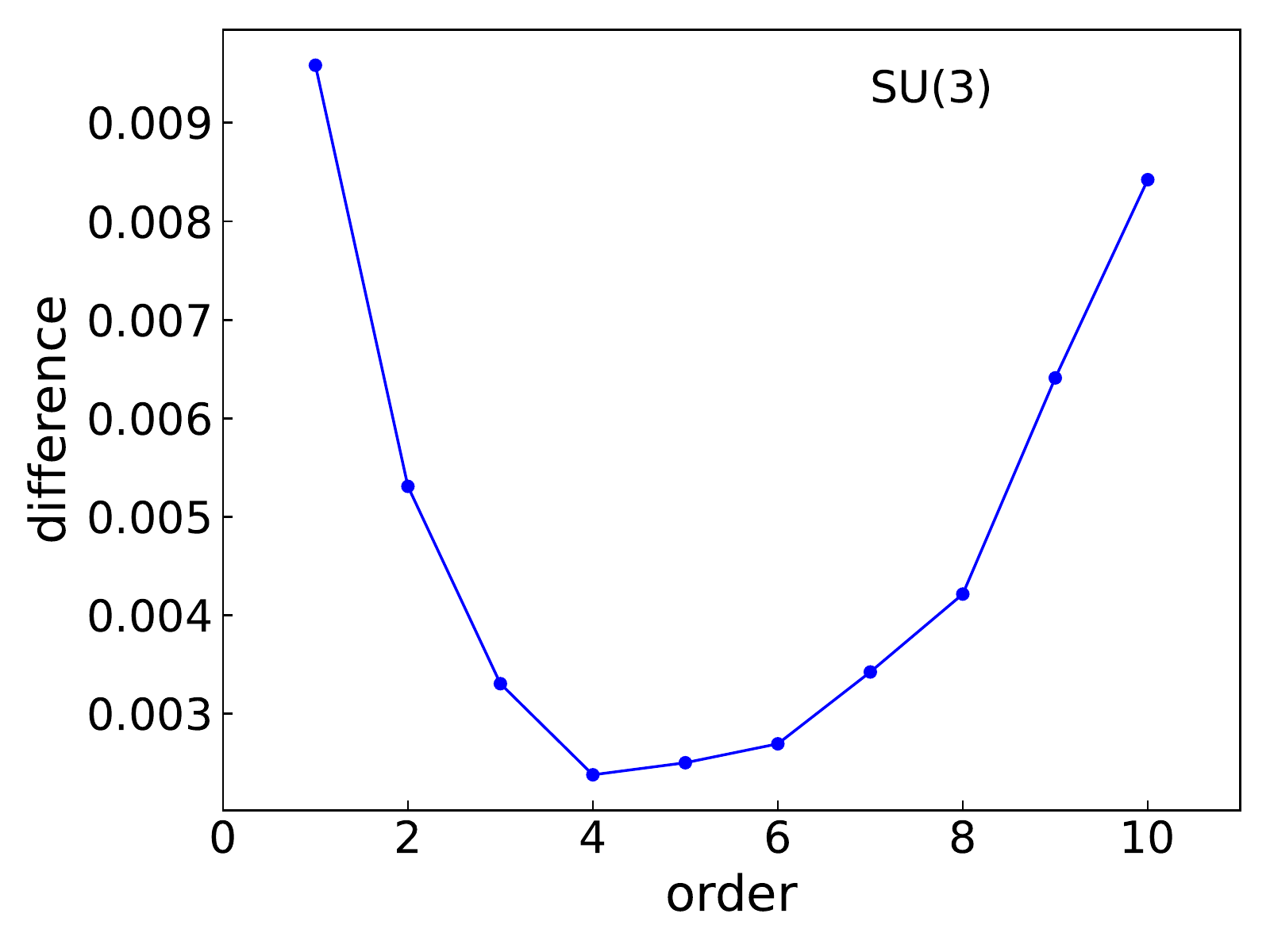}
\caption{The difference between the exact solution and 
the N$^n$LO calculations for SU(3) at $Z=0.84$ (left panel) and $Z=0.78$ (right panel). 
}
\label{fig:optimal_trancation}
\end{figure}

Usually, the partial sum of the asymptotic expansion 
(order by order) will first approach the true value, 
then it starts to diverge for larger orders. The perturbative solution is 
known to be good approximation up to the term of the order of the inverse of coupling constant.
Let us estimate what order in the perturbative series the 
expansion around $1/\kappa\approx 0$  is optimal 
(optimal truncation). By using the Stirling approximation, 
we make an rough estimate for the coefficients:
\begin{equation}
c_k \sim \frac{1}{k!}\left(\frac{a^2}{\kappa}\right)^k N_{4k}
 = \frac{\Gamma(2k+(N_c^2-1)/2)}{\Gamma((N_c^2-1)/2)}\frac{(-4x)^k}{k!}
 \sim (-16x)^k k! \,, 
\end{equation}
which grow factorially at large $k$.
This gives the order estimate of the optimal number of terms 
$k_\mathrm{opt}\sim 1/(16x)=Z^2/(16y)$~\cite{Marino:2012zq}.
When we substitute the value of $y=0.004$ at $Z=0.84$,
the optimal value is $k_\mathrm{opt}\approx11$.
For the value of $y=0.0057$ at $Z=0.78$,
the optimal value is $k_\mathrm{opt}\approx 6$.
In Fig.~\ref{fig:optimal_trancation}, we show the difference between 
the exact solution and the result of perturbation at $Z=0.84$ and $Z=0.78$
as a function of the order in perturbation theory in SU(3). 
The origin of the divergence are non-perturbative terms that do not contain in a Taylor expansion~\cite{Kazakov:1980rd,Flory:2012nk}. 

\section{Perturbation theory in the limit of large non-Gaussian fluctuations: the N$^n$LO order}\label{sec:z}


In this section, we consider an arbitrary order in perturbation theory
in the limit of large non-Gaussian fluctuations.
The integral for the evaluation of each order in this regime is given by
\begin{equation}\label{eq:Nn_smallZ}
N_n \equiv \frac{1}{\mathcal{N}_0}\int\prod_a d\rho_x^a\, \rho_x^n\, e^{-S_W}
=\frac{\int dr\, r^{N_c^2-2+n}\, e^{-S_r}}{\int dr\,r^{N_c^2-2}\, e^{-S_r}}\,,
\end{equation}
where $S_W=\int d^2\rho_x\, \rho_x^4/\kappa$
and $S_r= a^2\,r^4/\kappa$.
By using 
\begin{equation}
\int_0^\infty dx\, x^n\, \exp\left[-\frac{x^4}{\kappa}\right]
=\frac{1}{4}\Gamma\left(\frac{n+1}{4}\right)\kappa^{\frac{n+1}{4}}\,,
\end{equation}
and $\Gamma(z+1)=z\,\Gamma(z)$, the integral in Eq. (\ref{eq:Nn_smallZ}) 
yields: 
\begin{equation}\label{eq:Nn_smallZ_2}
N_n 
=\frac{(N_c^2-1)}{4}\frac{\Gamma((N_c^2-1+n)/4)}{\Gamma((N_c^2+3)/4)}
   \left(\frac{\kappa}{a^2}\right)^{n/4}
=\frac{(N_c^2+n-5)}{4}\frac{\kappa}{a^2}\, N_{n-4}\,,
\end{equation}
with the initial conditions $N_0=1$ and
\begin{equation}
N_2
=\frac{\Gamma((N_c^2+1)/4)\,(\kappa/a^2)^{(N_c^2+1)/4}}{\Gamma((N_c^2-1)/4)\,(\kappa/a^2)^{(N_c^2-1)/4}}
=\frac{(N_c^2-1)}{4}\frac{\Gamma((N_c^2+1)/4)}{\Gamma((N_c^2+3)/4)}\frac{\sqrt{\kappa}}{a}\,.
\end{equation}
By using Eq. (\ref{eq:Nn_smallZ_2})
one can calculate the two- and 
-four point function of color charges for an arbitrary order 
{$\mathcal{O}(Z^n)$} in perturbation theory:
\begin{eqnarray}
\langle \rho^a_{x}\rho^a_{x}\rangle
&=&\left(\sum_{k=0}^n\frac{1}{k!}(-x)^k N_{4k+2} \right)
  \left(\sum_{k=0}^n\frac{1}{k!}(-x)^k N_{4k}\right)^{-1} 
=(N_c^2-1)\frac{\sqrt\kappa}{a}\sum_{k=0}^{n}d_k\, w^k\,,
  \\
\langle \rho^a_{x}\rho^a_{x}\rho^b_{x}\rho^b_{x}\rangle 
&=&\left(\sum_{k=0}^n\frac{1}{k!}(-x)^k N_{4k+4} \right)
  \left(\sum_{k=0}^n\frac{1}{k!}(-x)^k N_{4k}\right)^{-1}
  =\frac{(N_c^2-1)}{4}\frac{\kappa}{a^2}\left(1-2\sum_{k=0}^{n-1}d_k\,w^{k+1}\right)\,,
\end{eqnarray}
where $x=Z\,a^2/2\,\bar\mu^2$ and $w=\sqrt{\kappa}\,x/a\,=\,Z\,a\,\sqrt{\kappa}/2\,\bar\mu^2$.
The coefficients $d_k$ are listed in the Appendix \ref{appendix:coefficients_dk}.
The requirement of matching the two-point function of color charges 
of the MV model yields the following constraint 
\begin{equation}
 2\sum_{k=0}^{n} d_k\, w^{k+1} = Z
 \label{eq:rcond}
\end{equation}
at the order $\mathcal{O}(Z^n)$.
Thus, by noting that $w^2=Z^2/4y$, the ratio of the four-point function of color charges 
in the non-Gaussian and Gaussian theories is given by
\begin{equation}\label{eq:ratioRho4_smallZ_asymptotic}
R_n = \frac{w^2}{(N_c^2+1)Z^2}\left(1 -2\sum_{k=0}^{n-1}d_k w^{k+1}\right)
\approx \frac{w^2/Z^2}{N_c^2+1}(1-Z)
=\frac{1-Z}{4(N_c^2+1)y}
\,,
\end{equation}
which is the same as Eq.~(\ref{eq:ratio1}) that is obtained by the expansion in $1/\kappa$.
In the $Z\to0$ limit, we have $y=d_0^2$ then, 
\begin{equation}
R_0 =
\frac{1}{4(N_c^2+1)d_0^2}=\frac{4\,\Gamma((N_c^2+3)/4)^2}{(N_c^2+1)\Gamma((N_c^2+1)/4)^2},
\end{equation}
which is consistent with our previous work~\cite{Giannini:2020xme}.
%
%
Lastly, at the order $\mathcal{O}(Z)$ (LO), $R_1$ is given by
\begin{equation}
R_1= \frac{(-d_0+D)^2}{4(N_c^2+1)d_1^2Z^2}
   \left(1+\frac{d_0}{d_1}(d_0-D)\right) , ~~~D=\sqrt{d_0^2+2d_1Z}\,,
\end{equation}
recovering the result presented in~\cite{Giannini:2020xme}.

\begin{figure}[htb]
\includegraphics[width=8.0cm]{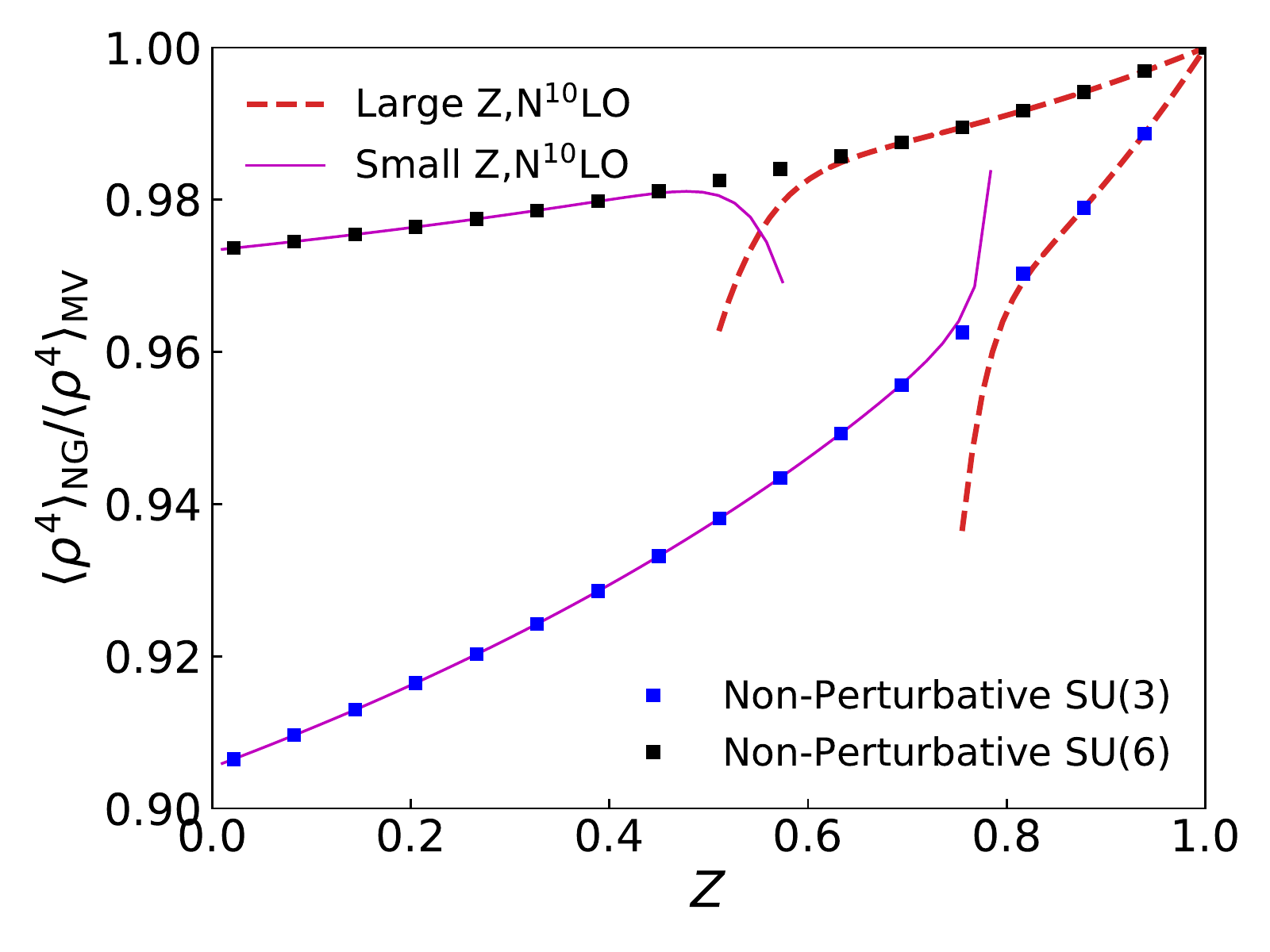}
\includegraphics[width=8.0cm]{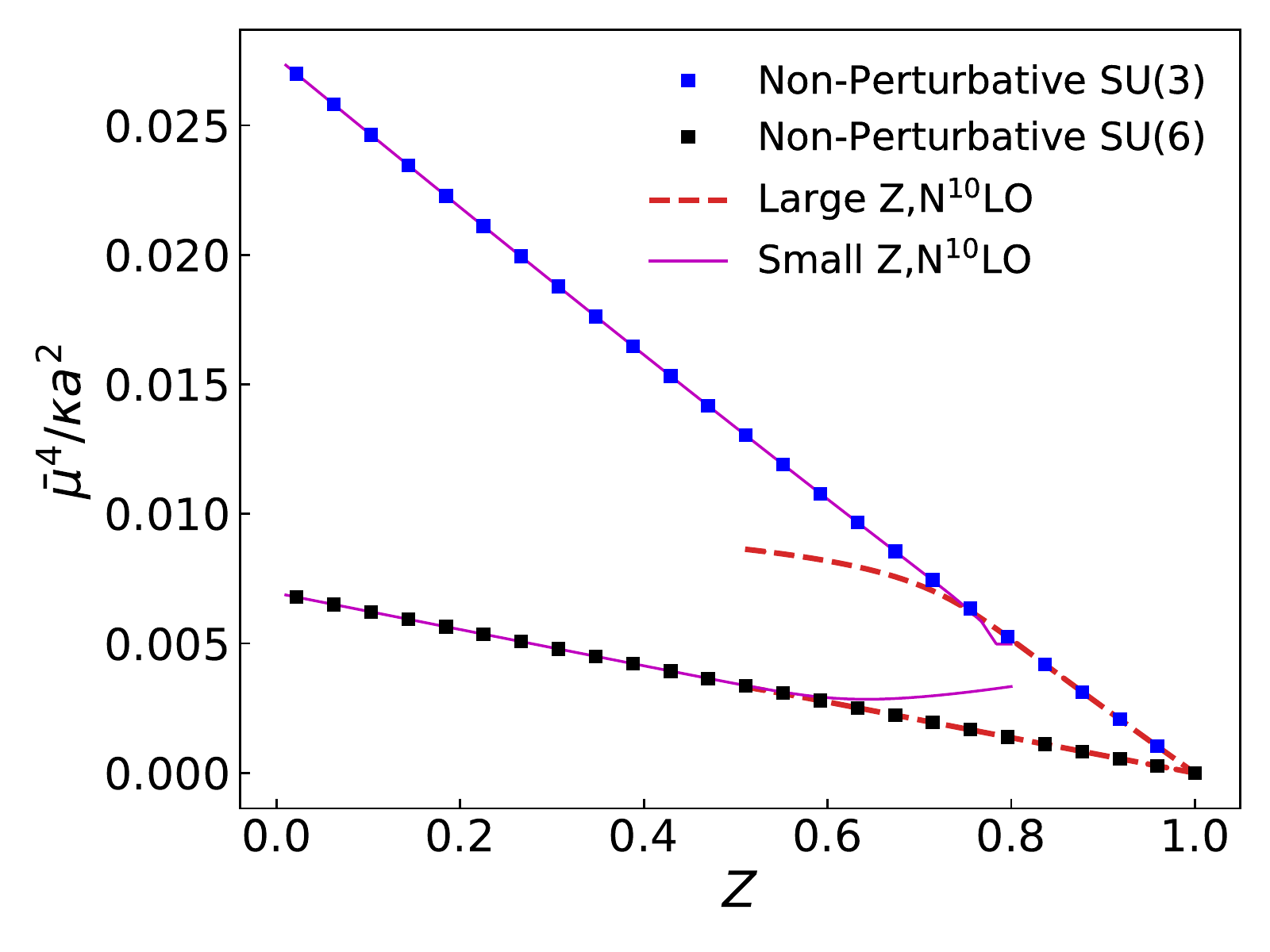}
\caption{Left panel: The ratio of the four-point function as a function of $Z$
for SU(3) and SU(6) group.
Right panel: $Z$ dependence of $y=\bar\mu^4/\kappa a^2$.
}
\label{fig:ratiosu7}
\end{figure}

Let us now compare results from both perturbative 
calculations worked out in this and previous sections 
to the full non-perturbative (numerical) calculation.
The left panel of Fig.~\ref{fig:ratiosu7} shows the ratio of the four-point function of color charges in the non-Gaussian and the Gaussian theories for SU(3) and SU(6)
from the perturbation in $1/\kappa\to0$ (dashed lines) and $Z\to0$ (solid lines), 
corresponding to small and large deviations form the MV model, respectively.
The perturbation in $Z$ shows a good agreement with the non-perturbative
calculation up to $Z=0.7$ for SU(3).
We showed in~\cite{Giannini:2020xme} that deviations 
from the MV model for $\langle \rho_x^a\rho_x^a\rho_x^b\rho_x^b \rangle$ 
when $Z\to0$ -- which is the region where one finds 
maximum deviation in both theories for this particular correlator -- 
become smaller as $N_c$ increases 
(with differences fully disappearing in the limit of large $N_c$).
Thus, by increasing $N_c$ we expect that the perturbation in $1/\kappa$ becomes 
better while the perturbation in $Z$ becomes worse.
Indeed, comparing the results for SU(3) and SU(6) confirms this expectation.
In the right panel of Fig.~\ref{fig:ratiosu7}, we also compare $Z$ dependence of 
$y=\bar\mu^4/\kappa a^2$ with the non-perturbative results obtained by 
numerical integration. 
The range of agreement between the different perturbative calculations 
and non-perturbative one for the $Z$ dependence of $y$ is similar to 
the range of agreement for the four-point function of color charges.

\section{The Borel-Pad\'e resummation}
\label{sec:BorelPade}

The previous sections showed that the perturbative series for the four-point function of color charges in both the limit of small and large non-Gaussian fluctuations diverges.
In this section, we employ the Borel-Pad\'e resummation method~\cite{Caliceti:2007ra}
to construct a convergent series for our two- and four-point functions of color charges.

The procedure of the Borel-Pad\'e resummation is the following.
Let $Z(y)=\sum_{n=0}^\infty c_n y^n$ be a divergent series,
where $c_n$ are the coefficients of the perturbative series and $y$ is 
the expansion parameter. 
As a first step, 
one calculates the Borel-transformed coefficients: $b_n=c_n/n!$
to remove the factorial growth of each coefficient,
which is a typical reason for the divergent series.
The coefficients $b_n$ define the Borel-transformed series:
$B(\tau)=\sum_{n=0}b_n \tau^n$.
In the Borel-Pad\'e summation, Pad\'e approximants are then 
used to approximate the Borel sum by a rational function:
\begin{equation}
 B_{L/M}(\tau)=\frac{\sum_{n=0}^L p_n \tau^n}{1+\sum_{n=1}^M q_n \tau^n},
\end{equation}
where $N=L+M$ gives the order of the Pad\'e approximant. 
The coefficients $q_n$ and $p_n$ are found 
by equating order by order the Taylor series of $B(\tau)$. 
Then one performs the Laplace transformation, 
\begin{equation}
 Z_{B,L/M}(y)=\int_0^\infty e^{-\tau} B_{L/M}(\tau y)d\tau\,,
\end{equation}
in order to reintroduce the contribution of the factorial 
factors removed in the first step.

As an example, let us consider the perturbative series
in the limit of small non-Gaussian fluctuations for $L=M=1$, in which
we solve Eq.~(\ref{eq:rz}) up to the order of $N=2$:
\begin{equation}
 Z= 1 - 4(N_c^2+1)y + 32(N_c^2+1)y^2\,.
\end{equation}
The [L/M] = [1/1] Pad\'e approximant is given by
\begin{equation}
 B_{[1/1]} = \frac{1-4N_c^2 \tau}{1+4\tau}\,.
\end{equation}
The value $y$ is obtained by solving
\begin{equation}
 Z = \int_0^\infty e^{-\tau}\frac{1-4N_c^2\tau y}{1+4\tau y}d\tau
\end{equation}
for a fixed value of $Z$ and the ratio of four-point 
functions of color charges is given by
\begin{equation}
 R_{[1/1]} = \int_0^\infty e^{-\tau}\frac{1+2(N_c^2+3)\tau y}{1+2(N_c^2+7)\tau y}d\tau\,.
\end{equation}
Higher-order approximants can be easily obtained 
in the Borel-Pad\'e resummation method, usually providing 
a better approximation for the divergent series (in case 
the series is Borel-Pad\'e summable). 

\begin{figure}[htb]
\includegraphics[width=8.0cm]{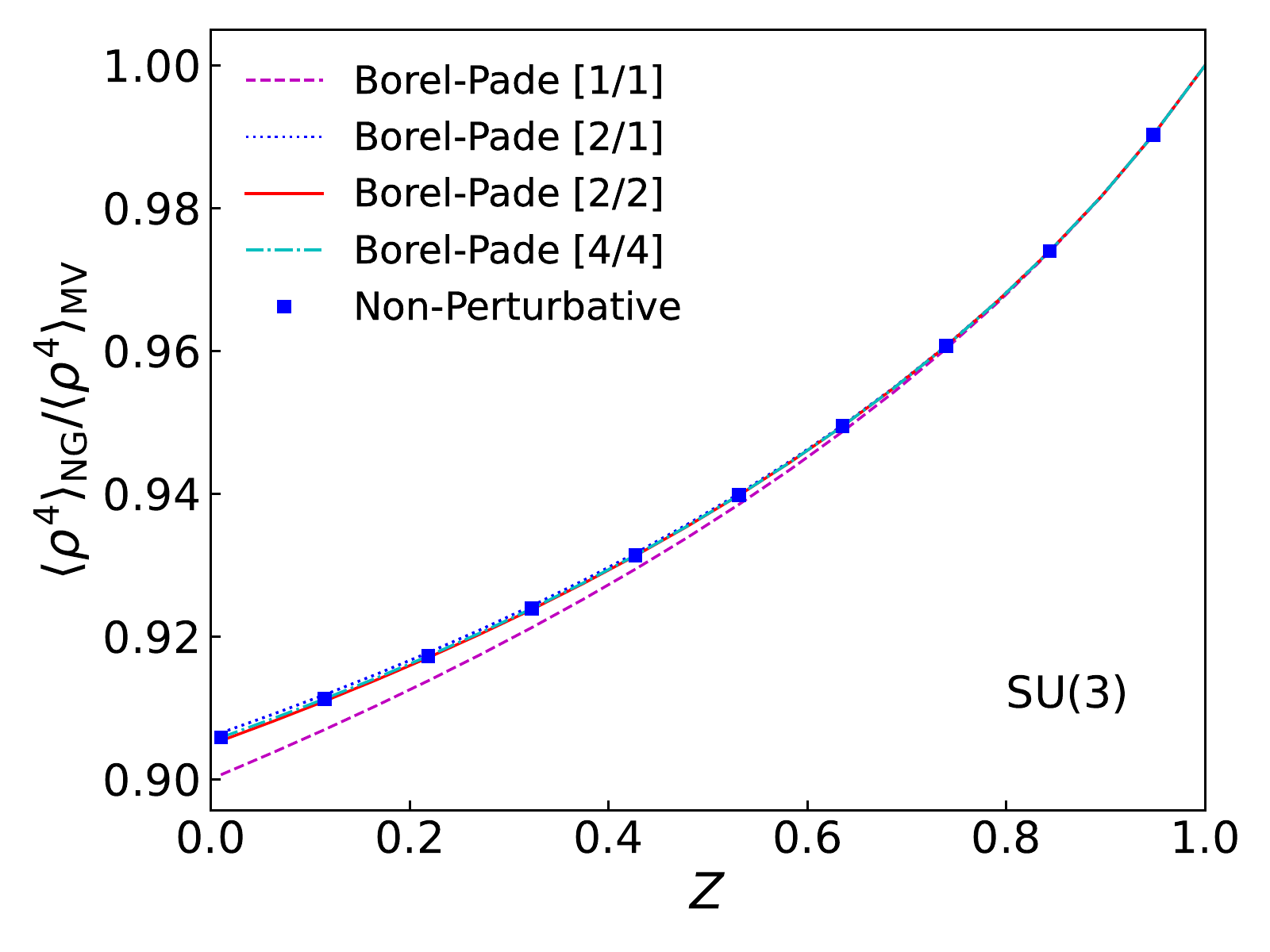}
\includegraphics[width=8.0cm]{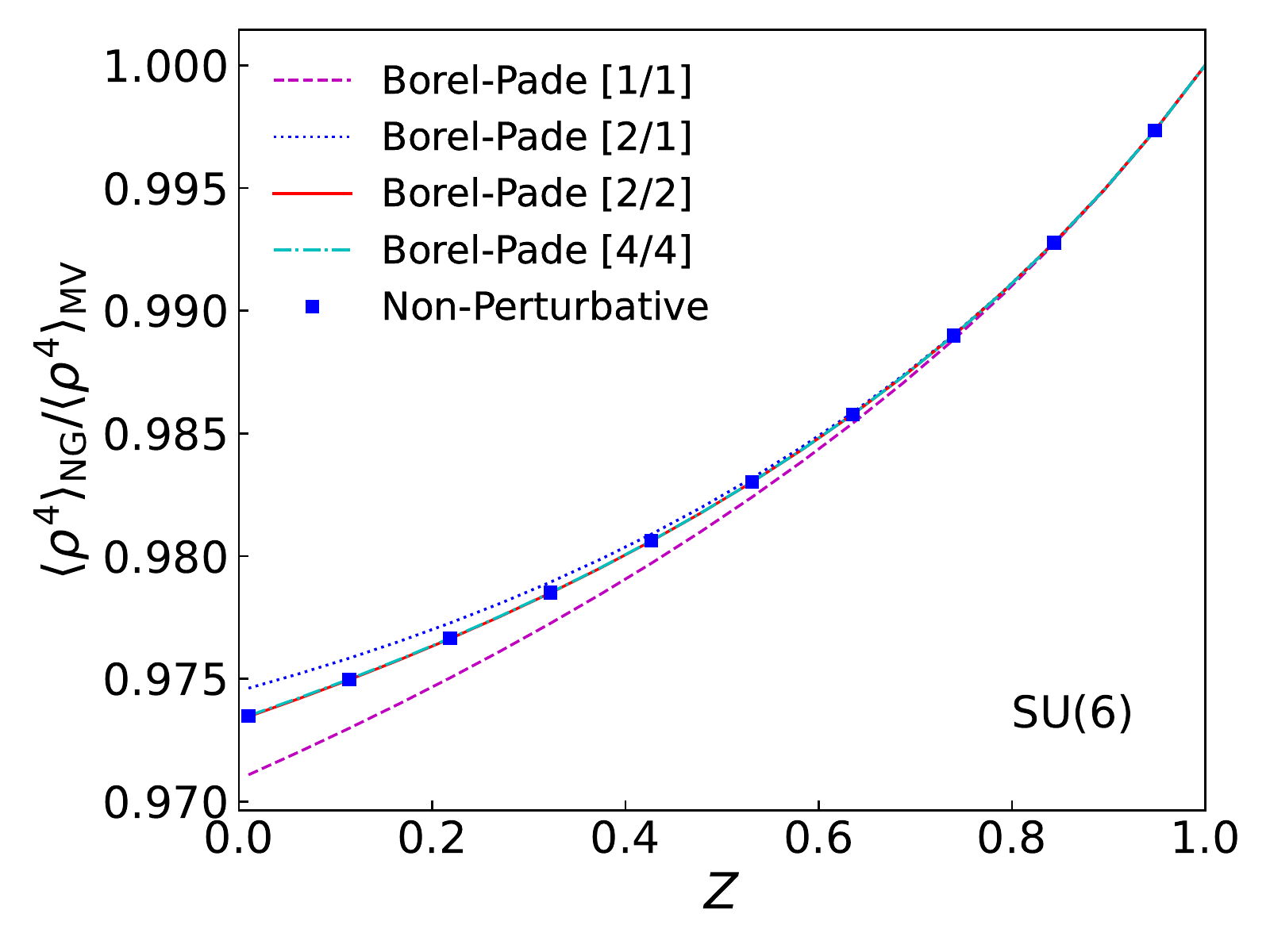}
\caption{The ratio of the four-point function as a function of $Z$
for SU(3) (left panel) and SU(6) group (right panel).
}
\label{fig:ratioBP}
\end{figure}

Figure \ref{fig:ratioBP} shows a comparison of Borel-Pad\'e 
approximants of different orders with the fully non-perturbative result
for the ratio of the four-point function of color charges
as a function of the renormalization factor, $Z$,
for SU(3) (left panel) and SU(6) (right panel).
As can be seen, the Borel-Pad\'e resummation of order [2/2] can reproduce in a good approximation the entire $Z$ dependence of the ratio 
for SU(3) and SU(6), even though we only utilize
information from the regime of small deviations from the MV model.
A possible reason for this fast convergence is due to the small values of $y$,
which are fixed by the condition Eq.~(\ref{eq:renom}) that the two-point functions 
of the non-Gaussian and the Gaussian theory are the same.
The condition gives the value of $y=d_0^2=0.027$ at $Z=0$ for SU(3)
as worked out in Sec.~\ref{sec:z}.

\section{Conclusions}
\label{sec:conclusion}

In this work, we considered the extension of the MV model by adding the first 
C-even correction and studied the structure 
of the perturbative series for the two- and four-point function of color charges
at N$^n$LO orders in perturbation theory both
in the limits of small and large non-Gaussian color charges fluctuations. 

Starting with the regime of small deviations from the MV 
model, we first extended previous results for these 
correlators to NLO and NNLO in $1/\kappa$ by 
explicitly calculating each Feynman diagram 
contributing at each one of these orders in perturbation. 
Next, a way to calculate these correlators 
at an arbitrary order, N$^n$LO, in $1/\kappa$ was 
provided. We found a simple expression for the  
four-point function of color charges.
Such a calculation allowed us to explicitly 
show that the resulting perturbative series is 
divergent; the optimal truncation for the perturbative 
series was estimated.
A similar procedure to calculate the two- and 
four-point function of color charges at N$^n$LO 
in the limit of large non-Gaussian color charges 
fluctuations has also been presented. 

Lastly, we showed that the perturbative series 
in $1/\kappa$ is Borel-Pad\'e summable, and one is 
able to recover the non-perturbative result by 
employing a Borel-Pad\'e approximant that only 
involves the first few terms of this series.

Extensions of the MV model 
of Gaussian color charges fluctuations may be relevant 
for describing small collision systems in the 
Color Glass Condensate framework.
This model may potentially be relevant and useful for the modelling of energy-momentum correlations in the initial state of heavy-ion collisions~\cite{Albacete:2018bbv},
and of eccentricities and their fluctuations~\cite{Dumitru:2012yr,Giacalone:2019kgg}.

\begin{acknowledgments}
We are grateful to Adrian Dumitru for his helpful comments, and careful reading of the manuscript.
Y. N. acknowledges the support by
the Grants-in-Aid for Scientific Research from JSPS (JP21K03577). 
A. V. G. acknowledges the support from FAPESP through the grants: 
17/05685-2, 17/14974-8, 18/24720-6 and 21/04924-9. 
\end{acknowledgments}

\appendix

\section{The two- and four-point function of color charges at NNLO in the limit of small non-Gaussian fluctuations}\label{appendix:nnlo}

The diagrams contributing for the two-point function at the
next-to-next-to-leading order (NNLO) 
in the limit of small non-Gaussian fluctuations are:
\begin{equation}
\begin{tikzpicture}[baseline=(h)]
\begin{feynman}[inline=(h)]
\vertex (a){\(y\)};
\vertex[below=2.0 cm of a](b){\(x\)};
\vertex[above=0.5cm of b](e);
\vertex[below=1.0cm of a](h);
\vertex[above=1.2cm of b](f);
\vertex[right=0.5cm of f](i);
\vertex[below=0.5cm of a](j);
\vertex[right=0.5cm of j](k);
\node at (-0.5,0.0) {(a)};
\node at (0.3,-0.5) {$z$};
\node at (-0.2,-0.7) {$s$};
\node at (-0.2,-1.6) {$w$};
\diagram*{
    (b)--(a),
   (f)--[out=-180,in=180,min distance=0.3cm](e),
   (f)--[out=0,in=0,min distance=0.3cm](e),
    (j)--[out=45,in=90,min distance=0.3cm](k),
    (j)--[out=-45,in=-90,min distance=0.3cm](k)
};
\end{feynman}
\end{tikzpicture}
= 3!\times2\times32(N_c^2+1)\times4(N_c^2+1)
\begin{tikzpicture}[baseline=(h)]
\begin{feynman}[inline=(h)]
\vertex (a){\(y\)};
\vertex[below=2.0 cm of a](b){\(x\)};
\vertex[above=0.5cm of b](e);
\vertex[below=1.0cm of a](h);
\vertex[above=1.5cm of b](f);
\vertex[right=0.5cm of f](i);
\vertex[below=1.0cm of a](j);
\vertex[right=0.3cm of j](k);
\vertex[right=0.5cm of k](l);
\node at (-0.5,0.0) {(b)};
\node at (0.55,-1.0) {$z$};
\node at (-0.2,-0.4) {$s$};
\node at (-0.2,-1.6) {$w$};
\diagram*{
    (b)--(a),
   (f)--[out=-180,in=180,min distance=0.3cm](e),
   (f)--[out=0,in=0,min distance=0.3cm](e),
    (k)--[out=45,in=90,min distance=0.3cm](l),
    (k)--[out=-45,in=-90,min distance=0.3cm](l)
};
\end{feynman}
\end{tikzpicture}
= 3!\times4^3\times6(N_c^2+1)^2
\end{equation}

\begin{equation}
\begin{tikzpicture}[baseline=(h)]
\begin{feynman}[inline=(h)]
\vertex (a){\(y\)};
\vertex[below=2.0 cm of a](b){\(x\)};
\vertex[right=0.45cm of a](x);
\vertex[below=0.5cm of x](e);
\vertex[below=1.5cm of x](f);
\vertex[below=1.0cm of a](h);

\vertex[right=0.45cm of a](y);
\vertex[below=0.5cm of y](i);
\vertex[below=1.5cm of y](j);
\node at (-0.5,0.0) {(c)};
\node at (-0.25,-1.0) {$z$};
\node at (0.45,-0.35) {$s$};
\node at (0.45,-1.65) {$w$};
\diagram*{
    (b)--(a),
   (f)--[out=-180,in=180,min distance=0.6cm](e),
   (f)--[out=0,in=0,min distance=0.6cm](e),
    (i)--[out=-130,in=130,min distance=0.01cm](j),
    (i)--[out=-30,in=30,min distance=0.01cm](j)
};
\end{feynman}
\end{tikzpicture}
= 3\times4^4(N_c^2+1)^2
\begin{tikzpicture}[baseline=(h)]
\begin{feynman}[inline=(h)]
\vertex (a){\(y\)};
\vertex[below=2.0 cm of a](b){\(x\)};
\vertex[below=1.0cm of a](h);
\vertex[below=0.6cm of a](e);
\vertex[below=1.4cm of a](f);
\vertex[right=0.6cm of h](g);

\vertex[right=0.45cm of a](y);
\vertex[below=0.5cm of y](i);
\vertex[below=1.5cm of y](j);
\node at (-0.5,0.0) {(d)};
\node at (0.80,-0.95) {$z$};
\node at (-0.25,-0.6) {$s$};
\node at (-0.25,-1.5) {$w$};
\diagram*{
    (b)--(a),(e)--(g),(f)--(g),
   (e)--[out=45,in=65,min distance=0.6cm](g),
   (f)--[out=-45,in=-45,min distance=0.6cm](g),
};
\end{feynman}
\end{tikzpicture}
= 3!\times4^3\times2(N_c^2+1)(N_c^2+7)
\end{equation}

\begin{equation}
\begin{tikzpicture}[baseline=(x)]
\begin{feynman}[inline=(x)]
\vertex (a){\(y\)};
\vertex[below=2.0 cm of a](b){\(x\)};
\vertex[above=0.5cm of b](e);
\vertex[right=0.5cm of e](h);
\vertex[above=1.5cm of b](f);
\vertex[right=0.5cm of f](i);
\vertex[below=1.0 cm of a ](x);
\vertex[right=0.5 cm of x ](y);
\node at (0.5,0.0) {(e)};
\node at (0.3,-1.5) {$z$};
\node at (0.3,-1.0) {$s$};
\node at (0.3,-0.5) {$w$};
\diagram*{
    (b)--(a),
    (e)--[out=45,in=90,min distance=0.3cm](h),
    (e)--[out=-45,in=-90,min distance=0.3cm](h),
    (f)--[out=45,in=90,min distance=0.3cm](i),
    (f)--[out=-45,in=-90,min distance=0.3cm](i),
    (x)--[out=45,in=90,min distance=0.3cm](y),
    (x)--[out=-45,in=-90,min distance=0.3cm](y)
};
\end{feynman}
\end{tikzpicture}
=3!\times[4(N_c^2 + 1)]^3
\begin{tikzpicture}[baseline=(h)]
\begin{feynman}[inline=(h)]
\vertex (a){\(y\)};
\vertex[below=2.0 cm of a](b){\(x\)};
\vertex[above=1.0cm of b](e);
\vertex[right=0.5cm of e](f);
\vertex[right=0.5cm of e](h);
\vertex[right=0.5cm of f](i);
\vertex[right=0.5cm of i](j);
\node at (0.5,0.0) {(f)};
\node at (0.25,-1.0) {$s$};
\node at (0.8,-1.0) {$w$};
\node at (1.3,-1.0) {$z$};
\diagram*{
    (b)--(a),
    (e)--[out=45,in=90,min distance=0.3cm](h),
    (e)--[out=-45,in=-90,min distance=0.3cm](h),
    (f)--[out=45,in=90,min distance=0.3cm](i),
    (f)--[out=-45,in=-90,min distance=0.3cm](i),
    (i)--[out=45,in=90,min distance=0.3cm](j),
    (i)--[out=-45,in=-90,min distance=0.3cm](j)
};
\end{feynman}
\end{tikzpicture}
=3!\times[4(N_c^2 + 1)]^3
\begin{tikzpicture}[baseline=(x)]
\begin{feynman}[inline=(x)]
\vertex (a){\(y\)};
\vertex[below=2.0 cm of a](b){\(x\)};
\vertex[above=1.3cm of b](e);
\vertex[right=0.5cm of e](f);
\vertex[right=0.5cm of e](h);
\vertex[right=0.5cm of f](i);
\vertex[above=0.7cm of b](j);
\vertex[right=0.5cm of j](k);
\vertex[below=1.0cm of a](x);
\node at (0.5,0.0) {(g)};
\node at (0.2,-1.3) {$z$};
\node at (0.2,-0.7) {$s$};
\node at (0.8,-0.70) {$w$};
\diagram*{
    (b)--(a),
    (e)--[out=45,in=90,min distance=0.3cm](h),
    (e)--[out=-45,in=-90,min distance=0.3cm](h),
    (f)--[out=45,in=90,min distance=0.3cm](i),
    (f)--[out=-45,in=-90,min distance=0.3cm](i),
    (j)--[out=45,in=90,min distance=0.3cm](k),
    (j)--[out=-45,in=-90,min distance=0.3cm](k)
};
\end{feynman}
\end{tikzpicture}
=3!\times2\times[4(N_c^2 + 1)]^3
\begin{tikzpicture}[baseline=(x)]
\begin{feynman}[inline=(x)]
\vertex (a){\(y\)};
\vertex[below=2.0 cm of a](b){\(x\)};
\vertex[above=0.8cm of b](s);
\vertex[right=0.3cm of s](e);
\coordinate (h) at ($(e)+({0.1+0.4/sqrt(50)},-{1.25/sqrt(4)})$);

\vertex[above=1.2cm of b](p);
\vertex[right=0.3cm of p](f);
\coordinate (i) at ($(f)+({0.10/sqrt(1)},{0.9/sqrt(3)})$);
\vertex[below=1.0 cm of a ](x);
\vertex[right=0.5 cm of x ](y);
\node at (0.5,0.0) {(h)};
\node at (0.4,-1.5) {$z$};
\node at (-0.1,-1.0) {$s$};
\node at (0.33,-0.50) {$w$};
\diagram*{
    (b)--(a),
    (e)--[out=-45,in=0,min distance=0.3cm](h),
    (e)--[out=-90,in=-180,min distance=0.3cm](h),

    (f)--[out=135,in=180,min distance=0.3cm](i),
    (f)--[out=45,in=0,min distance=0.3cm](i),

    (x)--[out=45,in=90,min distance=0.3cm](y),
    (x)--[out=-45,in=-90,min distance=0.3cm](y)

};
\end{feynman}
\end{tikzpicture}
=3\times2\times4^3(N_c^2 + 1)^3
\end{equation}

There are $3!=6$ permutations of the three vertices $z$-$w$-$s$
except the diagrams (c) and (h).
(c) and (h) have 3 distinguishable permutations for the three vertices
$z$-$w$-$s$. 
The sum of the color factors yield:
\begin{equation}
\frac{\text{sum}}{3!}=64(N_c^2+1)[(5(N_c^2+1)^2 + 12(N_c^2+1) + 2(N_c^2+7))]
=64(N_c^2+1)(5N_c^4 + 24N_c^2 + 31)\,,
\end{equation}
and the two-point function of color charges 
at NNLO order is given by:
\begin{eqnarray}
\langle \rho^a_x\rho^b_y\rangle
&=&\mu^2\frac{\delta^{ab}\delta_{xy}}{a^2}
\Bigg[
1-4\frac{\mu^4}{\kappa a^2}(N_c^2+1)
+32\frac{\mu^8}{\kappa^2a^4}
(N_c^2+1)(N_c^2+2)
-64\frac{\mu^{12}}{\kappa^3a^6}
(N_c^2+1)(5N_c^4+24N_c^2+31)\Bigg]\,.
\end{eqnarray}

Thus, renormalized $\mu$ is defined as
\begin{eqnarray}
\bar\mu^2
&=&\mu^2
\left[
1-4\frac{\mu^4}{\kappa a^2}(N_c^2+1)
+32\frac{\mu^8}{\kappa^2a^4}
(N_c^2+1)(N_c^2+2)
-64\frac{\mu^{12}}{\kappa^3a^6}
(N_c^2+1)(5N_c^4+24N_c^2+31)\right]\,.
\end{eqnarray}

The connected diagrams for four-point function at 
order $1/(3!\kappa^3)$ are:
\begin{equation}
\begin{tikzpicture}[baseline=(h)]
\begin{feynman}[inline=(h)]
\vertex (a){\(x\)};
\vertex[right=2 cm of a ](b){\(v\)};
\vertex[below=2 cm of b] (c){\(y\)};
\vertex[left=2 cm of c](d){\(u\)};
\path (a)--(c) coordinate[pos=0.25] (e);
\coordinate (h) at ($(e)+(-{0.6/sqrt(2)},-{0.6/sqrt(2)})$);
\coordinate (f) at ($(h)+(-{0.6/sqrt(2)},-{0.6/sqrt(2)})$);
\node at (0.4,-0.8) {$z$};
\node at (1.0,-1.3) {$w$};
\node at (-0.05,-1.25) {$s$};
\diagram*{
    (e)--[out=-180,in=135,min distance=0.3cm](h),
    (e)--[out=-90,in=-45,min distance=0.3cm](h),
    (c)--(a),
    (d)--[draw=white,double=black,very thick](b),
    (h)--[out=-180,in=135,min distance=0.3cm](f),
    (h)--[out=-90,in=-45,min distance=0.3cm](f)
};
\end{feynman}
\end{tikzpicture}
=3!\times4\times8[4(N_c^2+1)]^2
\begin{tikzpicture}[baseline=(h)]
\begin{feynman}[inline=(h)]
\vertex (a){\(x\)};
\vertex[right=2 cm of a ](b){\(v\)};
\vertex[below=2 cm of b] (c){\(y\)};
\vertex[left=2 cm of c](d){\(u\)};
\path (a)--(c) coordinate[pos=0.35] (e);
\path (a)--(c) coordinate[pos=0.10] (f);
\coordinate (h) at ($(e)+(-{0.6/sqrt(2)},-{0.6/sqrt(2)})$);
\coordinate (g) at ($(f)+(-{0.6/sqrt(2)},-{0.6/sqrt(2)})$);
\node at (0.5,-1.0) {$z$};
\node at (1.0,-1.3) {$w$};
\node at (0.1,-0.6) {$s$};
\diagram*{
    (e)--[out=-180,in=135,min distance=0.3cm](h),
    (e)--[out=-90,in=-45,min distance=0.3cm](h),
    (c)--(a),
    (d)--[draw=white,double=black,very thick](b),
    (f)--[out=-180,in=135,min distance=0.3cm](g),
    (f)--[out=-90,in=-45,min distance=0.3cm](g)
};
\end{feynman}
\end{tikzpicture}
=3!\times4\times8[4(N_c^2+1)]^2
\begin{tikzpicture}[baseline=(h)]
\begin{feynman}[inline=(h)]
\vertex (a){\(x\)};
\vertex[right=2 cm of a ](b){\(v\)};
\vertex[below=2 cm of b] (c){\(y\)};
\vertex[left=2 cm of c](d){\(u\)};
\path (a)--(c) coordinate[pos=0.25] (e);
\coordinate (h) at ($(e)+(-{0.6/sqrt(2)},-{0.6/sqrt(2)})$);
\path (b)--(d) coordinate[pos=0.25] (f);
\coordinate (g) at ($(f)+(+{0.6/sqrt(2)},-{0.6/sqrt(2)})$);
\node at (0.4,-0.8) {$z$};
\node at (1.0,-1.3) {$w$};
\node at (1.6,-0.8) {$s$};
\diagram*{
    (e)--[out=-180,in=135,min distance=0.3cm](h),
    (e)--[out=-90,in=-45,min distance=0.3cm](h),
    (c)--(a),
    (d)--[draw=white,double=black,very thick](b),
    (f)--[out=-75,in=-135,min distance=0.3cm](g),
    (f)--[out=0,in=45,min distance=0.3cm](g)
};
\end{feynman}
\end{tikzpicture}
=3!\times6\times8[4(N_c^2+1)]^2
\end{equation}

\begin{equation}
\begin{tikzpicture}[baseline=(h)]
\begin{feynman}[inline=(h)]
\vertex (a){\(x\)};
\vertex[right=2 cm of a ](b){\(v\)};
\vertex[below=2 cm of b] (c){\(y\)};
\vertex[left=2 cm of c](d){\(u\)};
\path (a)--(c) coordinate[pos=0.06] (e);
\path (a)--(c) coordinate[pos=0.35] (f);
\node at (0.3,-0.10) {$z$};
\node at (1.0,-1.3) {$w$};
\node at (0.6,-0.9) {$s$};
\diagram*{
    (c)--(a),
    (d)--[draw=white,double=black,very thick](b),
   (f)--[out=-180,in=270,min distance=0.4cm](e),
   (f)--[out=90,in=0,min distance=0.4cm](e)
};
\end{feynman}
\end{tikzpicture}
=3!\times4\times8\times32(N_c^2+1)
\end{equation}

\begin{equation}
\begin{tikzpicture}[baseline=(v)]
\begin{feynman}[inline=(v)]
\vertex (a){\(x\)};
\vertex[right=2 cm of a ](b){\(y\)};
\vertex[below=1.0 cm of a] (x);
\vertex[right=1.0 cm of x] (e);
\vertex[below=0.5 cm of e] (v);
\vertex[below=1.0 cm of e] (f);
\vertex[below=1.0 cm of f] (y);
\vertex[left=1.0 cm of y] (c){\(u\)};
\vertex[right=1.0 cm of y] (d){\(v\)};
\path (a)--(e) coordinate[pos=0.35] (g);
\coordinate (h) at ($(g)+(-{0.5/sqrt(2)},-{0.5/sqrt(2)})$);
\node at (0.25, -0.7) {$s$};
\node at (1.0,-1.8) {$z$};
\node at (1.0,-1.2) {$w$};
\diagram*{
   (a)--(e),(e)--(b),(c)--(f),(d)--(f),
   (f)--[out=-180,in=180,min distance=0.5cm](e),
   (f)--[out=0,in=0,min distance=0.5cm](e),
    (g)--[out=-180,in=135,min distance=0.3cm](h),
    (g)--[out=-90,in=-45,min distance=0.3cm](h)
};
\end{feynman}
\end{tikzpicture}
=3!\times4\times32(N_c^2+7)\times4(N_c^2+1)
\begin{tikzpicture}[baseline=(v)]
\begin{feynman}[inline=(v)]
\vertex (a){\(x\)};
\vertex[right=2 cm of a ](b){\(y\)};
\vertex[below=1.0 cm of a] (x);
\vertex[right=1.0 cm of x] (e);
\vertex[below=0.5 cm of e] (v);
\vertex[below=1.0 cm of e] (f);
\vertex[below=1.0 cm of f] (y);
\vertex[left=1.0 cm of y] (c){\(u\)};
\vertex[right=1.0 cm of y] (d){\(v\)};
\path (e)--(f) coordinate[pos=0.5] (i);
\coordinate (g) at ($(i)+(0.4,0)$);
\coordinate (h) at ($(g)+(0.5,0)$);
\node at (1.7, -1.5) {$s$};
\node at (1.0,-1.8) {$z$};
\node at (1.0,-1.2) {$w$};
\diagram*{
   (a)--(e),(e)--(b),(c)--(f),(d)--(f),
   (f)--[out=-180,in=180,min distance=0.5cm](e),
   (f)--[out=0,in=0,min distance=0.5cm](e),
    (g)--[out=-45,in=-90,min distance=0.5cm](h),
    (g)--[out=45,in=90,min distance=0.5cm](h)
};
\end{feynman}
\end{tikzpicture}
=3!\times2\times32(N_c^2+7)\times4(N_c^2+1)
\end{equation}

\begin{equation}
\begin{tikzpicture}[baseline=(v)]
\begin{feynman}[inline=(v)]
\vertex (a){\(x\)};
\vertex[right=1 cm of a ](b){\(y\)};
\vertex[below=0.5 cm of a] (x);
\vertex[right=0.5 cm of x] (e);
\vertex[below=0.5 cm of e] (v);
\vertex[below=0.5 cm of e] (f);
\vertex[below=0.5 cm of f] (g);
\vertex[below=0.5 cm of g] (y);
\vertex[left=0.5 cm of y] (c){\(u\)};
\vertex[right=0.5 cm of y] (d){\(v\)};
\node at (0.5,-0.3) {$z$};
\node at (1.0,-1.0) {$w$};
\node at (0.5,-1.8) {$s$};
\diagram*{
   (a)--(e),(e)--(b),(c)--(g),(d)--(g),
   (f)--[out=-180,in=180,min distance=0.3cm](e),
   (f)--[out=0,in=0,min distance=0.3cm](e),
   (f)--[out=-180,in=180,min distance=0.3cm](g),
   (f)--[out=0,in=0,min distance=0.3cm](g)
};
\end{feynman}
\end{tikzpicture}
=3!\times128(N_c^4+4N_c^2 + 15)
~~~
\begin{tikzpicture}[baseline=(v)]
\begin{feynman}[inline=(v)]
\vertex (a){\(x\)};
\vertex[right=1 cm of a ](b){\(y\)};
\vertex[below=0.5 cm of a] (x);
\vertex[right=0.5 cm of x] (e);
\vertex[below=0.5 cm of e] (v);
\vertex[below=1.0 cm of e] (f);
\vertex[below=0.5 cm of e] (h);
\vertex[left=0.3 cm of h] (i);
\vertex[right=0.3 cm of h] (j);

\vertex[below=0.7 cm of h] (k);
\vertex[left=0.5 cm of k] (c){\(u\)};
\vertex[right=0.5 cm of k] (d){\(v\)};
\node at (0.5,-0.3) {$z$};
\node at (1.0,-1.0) {$w$};
\node at (0.0,-1.0) {$s$};
\diagram*{
   (i)--(j),
   (a)--(e),(e)--(b),(c)--(i),(d)--(j),
   (f)--[out=-180,in=180,min distance=0.3cm](e),
   (f)--[out=0,in=0,min distance=0.3cm](e),
};
\end{feynman}
\end{tikzpicture}
=3!\times2\times128(10N_c^2 + 34)
\end{equation}
We note that these connected graphs are also obtained by cutting the graphs of
the two-point function.
\begin{eqnarray}
\frac{\mathrm{sum}}{3!\kappa^3} &=&
128[ 16(N_c^2+1)^2 + 8(N_c^2+1) + 6(N_c^2+1)(N_c^2+7) + N_c^4 + 4N_c^2 + 15 +
2(10N_c^2+34)]\\
&=& 128( 21N_c^4 + 108N_c^2 + 147)\,.
\end{eqnarray}

\begin{equation}
\frac{\text{total sum}}{3!\kappa^3} =
128( 7N_c^6 + 37N_c^4 + 65N_c^2 + 35 + 21N_c^4 + 108N_c^2 + 147 )
= 128(7N_c^6 + 58N_c^4 + 173N_c^2 + 182)
\end{equation}

The renormalized expression at $1/\kappa^3$ is obtained as
\begin{eqnarray}
A_\mathrm{con} &=& \frac{\mu^4}{a^4}
 \left[
   -8\frac{\mu^{4}}{\kappa a^2}
  +32\frac{\mu^8}{\kappa^2a^4}(5N_c^2+11)
  -128\frac{\mu^{12}}{\kappa^3a^6}(21N_c^4+108N_c^2 + 147)
  \right]\\
&=& \frac{\bar\mu^4}{a^4}
 \left[
   -8\frac{\bar\mu^{4}}{\kappa a^2}
  +32\frac{\bar\mu^8}{\kappa^2a^4}(N_c^2+7)
  -128\frac{\bar\mu^{12}}{\kappa^3a^6}(N_c^4 + 24N_c^2 + 83)
  \right]
\end{eqnarray}

\section{Gaussian integral for $n$-th order}\label{appendix:GaussianIntegral}

The two- and four-point functions can be obtained by 
means of one-dimensional Gaussian integrations. We define $N_n$ as
\begin{equation}
N_n =\frac{1}{\mathcal{N}_0}\int \prod_a d\rho_x^a\, \rho_x^n\, e^{-S_G}\,.
\end{equation}
Using the following Gaussian integral
\begin{equation}
n_n =\int_0^{\infty} dx\, x^n
\exp\left[-\frac{a^2x^2}{2\mu^2}\right]=\frac{1}{2}\,
\Gamma\left(\frac{n+1}{2}\right)\left(\frac{\sqrt{2}\mu}{a}\right)^{n+1}\,.
\end{equation}
The normalization is obtained as
\begin{eqnarray}
n_0 &=& 
\frac{\sqrt{\pi}}{2}\left(\frac{\sqrt{2}\,\mu}{a}\right)\,,\\
\mathcal{N}_0 &=&\int \prod_a d\rho_x^a\,e^{-\rho_x^2\,a^2/(2\,\mu^2)}=n_0^{N_c^2-1}\,.
\end{eqnarray}
Then, for even $n$,
\begin{align}
N_n &=\frac{1}{\mathcal{N}_0}\int \prod_a d\rho_x^a\, \rho_x^n\, e^{-S_G}
=\frac{\int dr\, r^{N_c^2-2+n}\, e^{-a^2\,r^2/(2\,\mu^2)}}{\int dr\, r^{N_c^2-2}\, 
	e^{-a^2r^2/(2\mu^2)}}
=\frac{\Gamma(\frac{(N_c^2-1+n)}{2})\left(\frac{\sqrt{2}\,\mu}{a}\right)^{N_c^2-1+n}}
      {\Gamma(\frac{(N_c^2-1)}{2})\left(\frac{\sqrt{2}\,\mu}{a}\right)^{N_c^2-1}}\nonumber\\
&=\prod_{k=1}^{n/2}(N_c^2-1+n-2k)\left(\frac{\mu}{a}\right)^n
=(N_c^2-3+n)N_{n-2}\frac{\mu^2}{a^2}\,,
\end{align}
where we have used 
$\Gamma(z+1)=z\,\Gamma(z)$
to get
\begin{equation}
\Gamma\left(\frac{N_c^2-1+n}{2}\right)=\frac{N_c^2-1+n-2}{2}\,
\Gamma\left(\frac{N_c^2-1+n-2}{2}\right)\,.
\end{equation}
The first few terms are given by
\begin{align}
N_2 &= (N_c^2-1)\left(\frac{\mu}{a}\right)^2,~~~
N_4 = (N_c^2+1)\,N_2\,\left(\frac{\mu}{a}\right)^2,~~~
N_6 = (N_c^2+3)\,N_4\,\left(\frac{\mu}{a}\right)^2,\\
N_8 &= (N_c^2+5)\,N_6\,\left(\frac{\mu}{a}\right)^2,~~
N_{10} = (N_c^2+7)\,N_8\,\left(\frac{\mu}{a}\right)^{2}.
\end{align}

The four-point function in the NLO may be obtained as
\begin{align}
\langle \rho^4_x\rangle 
&=\frac{\int dr\, r^{N_c^2-2}\, r^4\, e^{-a^2r^2/(2\mu^2)}[1-\frac{a^2}{\kappa}r^4+\frac{a^4}{2\kappa^2}r^8]}
       {\int dr\,
       r^{N_c^2-2}\,e^{-a^2r^2/(2\mu^2)}[1-\frac{a^2}{\kappa}r^4+\frac{a^4}{2\kappa^2}r^8]}\nonumber \\
&=\left( N_4-\frac{a^2}{\kappa_4}N_8 + \frac{a^4}{2\kappa_4^2}N_{12}\right)
             \left(1-\frac{a^2}{\kappa_4}N_4 + \frac{a^4}{2\kappa_4^2}N_{8}\right)^{-1}\nonumber \\
&=  N_4-\frac{a^2}{\kappa_4}\left(N_8 - N_4^2 \right)
   +\frac{a^4}{2\kappa_4^2} \left(
    N_{12} - 3N_4N_{8} + 2N_4^3
	 \right)\,. \label{eq:NLOtermGaussianIntegration}
\end{align}
The last term in the above equation  
is evaluated as
\begin{eqnarray}
  \frac{1}{2} \left(
    N_{12} - 3N_4N_{8} + 2N_4^3
    \right)
&=&
   \frac{\mu^8}{2a^8}N_4 
   \left[
    (9)(7)(5)(3) - 3(5)(3)(1)(-1) + 2(N_c^4-1)^2 
	 \right] \nonumber \\
&=&  (N_c^4-1) \frac{\mu^{12}}{a^{12}} 
    16\left(
     5N_c^4 + 24N_c^2 + 31
	 \right)\label{eq:NLOtermGaussianIntegration2}
\end{eqnarray}
where $(n)\equiv (N_c^2+n)$. 
Thus, we get the same result shown in section \ref{sec:NLO_largeZ} after using 
Eq. (\ref{eq:NLOtermGaussianIntegration2}) into  
Eq. (\ref{eq:NLOtermGaussianIntegration}) and 
factoring out $(N_c^4-1)\mu^4/a^4$, which is common 
to all terms.

\section{Coefficients of the perturbative series for small non-Gaussian color charge fluctuations}\label{appendix:coefficients_ak_c_ck}

We list the coefficients $c$, $a_k$, and $c_k$ appearing in the 
perturbative series for the two- and four-point function 
of color charges in the limit of small non-Gaussian color charge 
fluctuations.
The ten first coefficients of the series before applying the 
renormalization procedure are:
\begin{eqnarray}
	c &=& 4(N_c^2 + 1)\,, \\
	a_0 &=& 1\,, \\
	a_1&=&-8(N_c^2 + 2)\,, \\
	a_2&=&16(5\,N_c^4 + 24\,N_c^2 + 31)\,, \\
	a_3&=&-128(7\,N_c^6 + 58\,N_c^4 + 173\,N_c^2 + 182)\,, \\
	a_4&=&512(21\,N_c^8 + 260\,N_c^6 + 1306\,N_c^4 + 3100\,N_c^2 + 2873)\,, \\
	a_5&=&-4096(33\,N_c^{10}+ 562\,N_c^8 + 4146\,N_c^6 + 16312\,N_c^4 + 33621\,N_c^2 + 28486)\,,\\
	a_6&=&4096(429\,N_c^{12} + 9520\,N_c^{10} + 95377\,N_c^8 + 544768\,N_c^6 + 1841863\,N_c^4
	+ 3437392\,N_c^2 + 2719291)\,,\\
	a_7 &=& -32768(715\,N_c^{14} + 19898\,N_c^{12} + 257207\,N_c^{10} + 1977490\,N_c^8
	+ 9630017\,N_c^6 + 29269870N_c^4 
	\nonumber\\ &+& 50652541\,N_c^2 + 37921862)\,, \\
	a_8 &=& 131072(2431\,N_c^{16} + 82452\,N_c^{14} + 1326100\,N_c^{12} + 13062468\,N_c^{10}\nonumber\\
	&+& 85096170\,N_c^8 + 370462908\,N_c^6 + 1038627956\,N_c^4 + 1692065772\,N_c^2 +
	1210080143)\,,\\
	a_9 &=& -1048576 (4199\,N_c^{18} + 169766\,N_c^{16} + 3306284\,N_c^{14} + 40291196\,N_c^{12}
	+ 334606770\,N_c^{10} \nonumber\\
	&+& 1939973736\,N_c^8 
	+ 7757897212\,N_c^6 + 20392359076\,N_c^4 + 31601443135\,N_c^2 + 21735270226)\,, \\
	a_{10} &=& 2097152(29393\,N_c^{20} + 1391720\,N_c^{18} + 32136215\,N_c^{16} + 471973040\,N_c^{14} 
	+ 4829037610\,N_c^{12} \nonumber\\
	&+& 35560069248\,N_c^{10} + 188741774926\,N_c^8 + 705359761744\,N_c^6 
	+ 1757875371253\,N_c^4 
	\nonumber\\
	&+& 2610819668248\,N_c^2 + 1735955801003) \,.
\end{eqnarray}

The ten first coefficients for the renormalized perturbative 
series are listed next.
\begin{eqnarray}
	c_0 &=& 1\,, \\
	c_1 &=& -8\,, \\
	c_2 &=& 32\,(N_c^2 + 7)\,, \\
	c_3 &=& -128\,(N_c^4 + 24\,N_c^2 + 83)\,, \\
	c_4 &=& 512\,(N_c^6 + 55\,N_c^4 + 571\,N_c^2 + 1357)\,, \\
	c_5 &=& -2048\,(N_c^8 + 104\,N_c^6  + 2266\,N_c^4  + 14976\,N_c^2 + 27933)\,, \\
	c_6 &=&  8192\,(N_c^{10} + 175\,N_c^8 + 6770\,N_c^6 + 88886\,N_c^4 + 438621\,N_c^2 + 688971)\,, \\
	c_7 &=& -32768\,(N_c^{12} + 272\,N_c^{10} + 16885\,N_c^{8} + 377600\,N_c^6
	+ 3564163\,N_c^4 + 14301296\,N_c^2 + 19746759)\,, \\
	c_8 &=& 131072\,(N_c^{14} + 399\,N_c^{12} + 37093\,N_c^{10}+ 1288331\,N_c^8 + 20104579\,N_c^6 
	+ 150092653\,N_c^4\,, \\
	&+&  515838295\,N_c^2 + 644057785 \nonumber\nonumber \\
	c_9 &=& -524288\,(N_c^{16} + 560\,N_c^{14} + 74116\,N_c^{12} + 3754624\,N_c^{10} + 88827950\,N_c^8 
	+ 1072010352\,N_c^6 \nonumber\nonumber \\
	&+& 6711810900\,N_c^4 + 20446471008\,N_c^2  + 23543136377 )\\
	c_{10} &=& 2097152\,(N_c^{18} + 759\,N_c^{16} + 137556\,N_c^{14}
	+ 9706268\,N_c^{12}
	+ 328292526\,N_c^{10}
	+ 5895609522\,N_c^8 
	\nonumber \\
	&+& 58633472228\,N_c^6 + 320047345068\,N_c^4 + 885046782489\,N_c^2 + 953277309583  )
\end{eqnarray}

\section{Coefficients of the perturbative series for large non-Gaussian color charge fluctuations}\label{appendix:coefficients_dk}

We list the coefficients $d_k$ appearing in the perturbative 
series for the two- and four-point function of color 
charges in the limit of large non-Gaussian fluctuations.
\begin{eqnarray}
d_0&=& \frac{1}{4}\frac{\Gamma((N_c^2+1)/4)}{\Gamma((N_c^2+3)/4)},\\
d_1 &=&d_0^2(N_c^2-1) - \frac{1}{4}, \\
d_2 &=&  d_0^3(N_c^2-1)^2 -\frac{d_0}{4}(N_c^2-2),\\
d_3&=& d_0^4(N_c^2-1)^3 -\frac{d_0^2}{3}(N_c^2-2)(N_c^2-1) +\frac{1}{48}(N_c^2-3),\\
d_4&=& d_0^5(N_c^2-1)^4 -\frac{5d_0^3}{12}(N_c^2-2)(N_c^2-1)^2
    +\frac{d_0}{96}(2N_c^2-5)(2N_c^2-3),\\
d_5&=& d_0^6(N_c^2-1)^5 -\frac{d_0^4}{2}(N_c^2-2)(N_c^2-1)^3
  +\frac{d_0^2}{240}(N_c^2-1)(17N_c^4-68N_c^2+65)
  -\frac{1}{480}(N_c^2-3)(N_c^2-2)\,, \\
d_6&=& d_0^7(N_c^2-1)^6 -\frac{7d_0^5}{12}(N_c^2-2)(N_c^2-1)^4
  +\frac{7d_0^3}{1440}(N_c^2-1)^2(22N_c^4-88N_c^2+85)\nonumber \\
  &-&\frac{d_0}{5760}(N_c^2-2)(34N_c^4-136N_c^2 + 117),\\
d_7&=& d_0^8(N_c^2-1)^7 -\frac{2d_0^6}{3}(N_c^2-2)(N_c^2-1)^5
  +\frac{d_0^4}{60}(N_c^2-1)^3(3N_c^2-7)(3N_c^2-5)\nonumber \\
  &-&\frac{d_0^2}{2520} (N_c^2-2)(N_c^2-1)(31N_c^4-124N_c^2 + 111)
   +\frac{1}{80640}(N_c^2-3)(17N_c^4-68N_c^2+63),\\
d_8&=& d_0^9(N_c^2-1)^8 -\frac{3d_0^7}{4}(N_c^2-2)(N_c^2-1)^6
  +\frac{d_0^5}{160}(N_c^2-1)^4(32N_c^4-128N_c^2 + 125)\nonumber \\
  &-&\frac{11d_0^3}{40320}(N_c^2-2)(N_c^2-1)^2(80N_c^4-320N_c^2 + 293)\nonumber \\
  &+& \frac{d_0}{645120}(496N_c^8-3968N_c^6+11468N_c^4 - 14128N_c^2 + 6237),\\
d_9&=& d_0^{10}(N_c^2-1)^9 -\frac{5d_0^8}{6}(N_c^2-2)(N_c^2-1)^7
  +\frac{d_0^6}{144}(N_c^2-1)^5(37N_c^4-148N_c^2 + 145)\nonumber \\
  &-&\frac{d_0^4}{1512}(N_c^2-2)(N_c^2-1)^3(53N_c^4-212N_c^2 + 197)\nonumber\\
  &+& \frac{d_0^2}{362880}(N_c^2-1)(691N_c^8-5528N_c^6+16106N_c^4 - 20200N_c^2 + 9180)
  \nonumber\\
  &-& \frac{1}{1451520}(N_c^2-3)(N_c^2-2)(31N_c^4-124N_c^2+105),\\
d_{10}&=& d_0^{11}(N_c^2-1)^{10} -\frac{11d_0^9}{12}(N_c^2-2)(N_c^2-1)^8
  +\frac{11d_0^7}{480}(N_c^2-1)^6(14N_c^4-56N_c^2 + 55) \nonumber\\
  &-&\frac{11d_0^5}{120960}(N_c^2-2)(N_c^2-1)^4(578N_c^4-2312N_c^2 + 2171)\nonumber \\
  &+& \frac{11d_0^3}{29030400}(N_c^2-1)^2(10256N_c^8-82048N_c^6+240256N_c^4 - 304640N_c^2 + 140895)\nonumber\\
  &-& \frac{d_0}{116121600}(N_c^2-2)((11056 N_c^8 - 88448 N_c^6  + 252064 N_c^4
  - 300672 N_c^2 + 126945)\,, \\
%
\end{eqnarray}

\end{document}